# Ammonium salts are a reservoir of nitrogen on a cometary nucleus and possibly on some asteroids

Olivier Poch[1]*, Istiqomah Istiqomah[1], Eric Quirico[1], Pierre Beck[1,2], Bernard Schmitt[1], Patrice Theulé[3], Alexandre Faure[1], Pierre Hily-Blant[1], Lydie Bonal[1], Andrea Raponi[4], Mauro Ciarniello[4], Batiste Rousseau[1]†, Sandra Potin[1], Olivier Brissaud[1], Laurène Flandinet[1], Gianrico Filacchione[4], Antoine Pommerol[5], Nicolas Thomas[5], David Kappel[6,7], Vito Mennella[8], Lyuba Moroz[7], Vassilissa Vinogradoff[9], Gabriele Arnold[7], Stéphane Erard[10], Dominique Bockelée-Morvan[10], Cédric Leyrat[10], Fabrizio Capaccioni[4], Maria Cristina De Sanctis[4], Andrea Longobardo[4,11], Francesca Mancarella[12], Ernesto Palomba[4], Federico Tosi[4]

**Affiliations:**
[1]Univ. Grenoble Alpes, Centre National de la Recherche Scientifique (CNRS), Institut de Planétologie et d'Astrophysique de Grenoble (IPAG), 38000 Grenoble, France
[2]Institut Universitaire de France (IUF), Paris, France
[3]Aix Marseille Univ, CNRS, Centre National d'Etudes Spatiales (CNES), Laboratoire d'Astrophysique de Marseille (LAM), Marseille, France
[4]Istituto di Astrofisica e Planetologia Spaziali (IAPS), Istituto Nazionale di Astrofisica (INAF), 00133 Rome, Italy
[5]Physikalisches Institut, Sidlerstr. 5, University of Bern, CH-3012 Bern, Switzerland
[6]Institute of Physics and Astronomy, University of Potsdam, 14476 Potsdam, Germany
[7]Institute for Planetary Research, German Aerospace Center (DLR), 12489 Berlin, Germany
[8]Istituto Nazionale di Astrofisica (INAF) – Osservatorio Astronomico di Capodimonte, Napoli, Italy
[9]Centre National de la Recherche Scientifique (CNRS), Aix-Marseille Université, Laboratoire Physique des Interactions Ioniques et Moléculaires (PIIM), Unité Mixte de Recherche (UMR) CNRS 7345, 13397 Marseille, France
[10]Laboratoire d'Etudes Spatiales et d'Instrumentation en Astrophysique (LESIA), Observatoire de Paris, Université Paris Sciences et Lettres (PSL), CNRS, Sorbonne Université, Université de Paris, 92195 Meudon, France
[11]Dipartimento di Scienze e Tecnologie (DIST), Università Parthenope, 80143 Napoli, Italy
[12]Dipartimento di Matematica e Fisica "E. De Giorgi", Università del Salento, Lecce, Italy
*Correspondence to: olivier.poch@univ-grenoble-alpes.fr
†Current address: Istituto di Astrofisica e Planetologia Spaziali (IAPS), Istituto Nazionale di Astrofisica (INAF), 00133 Rome, Italy

**Abstract:** The measured nitrogen-to-carbon ratio in comets is lower than for the Sun, a discrepancy which could be alleviated if there is an unknown reservoir of nitrogen in comets. The nucleus of comet 67P/Churyumov-Gerasimenko exhibits an unidentified broad spectral reflectance feature around 3.2 μm, which is ubiquitous across its surface. Based on laboratory experiments, we attribute this absorption band to ammonium salts mixed with dust on the surface. The depth of the band indicates that semi-volatile ammonium salts are a substantial reservoir of nitrogen in the comet, potentially dominating over refractory organic matter and more volatile species. Similar absorption features appear in the spectra of some asteroids, implying a compositional link between asteroids, comets and the parent interstellar cloud.

**Main Text:**

The composition of comets and asteroids can be investigated from the light scattered by their surfaces. For objects where the visible to near-infrared wavelength range shows no, or only weak, spectral features, analysis of the 3-μm region (between roughly 2.4 and 3.6 μm) can be used to investigate volatile and organic compounds present on their surfaces (*1*). The Visible, InfraRed and Thermal Imaging Spectrometer, Mapping Channel (VIRTIS-M) instrument (*2*) on the Rosetta spacecraft observed the nucleus of comet 67P/Churyumov-Gerasimenko (hereafter 67P) in the spectral range 0.2 to 5.1 μm (*3*). The surface imaged by VIRTIS-M appears almost spectrally uniform (*4*), characterized by a very low reflectance (geometric albedo of 6% at 0.55 μm (*5*)), positive (i.e. red) visible and infrared spectral slopes, and a broad absorption feature from 2.8 to 3.6 μm, centered at 3.2 μm (*3, 4*).

This absorption band, which has not been detected on other comets, is observed on all types of surface terrains, and was persistently observed from August 2014 when comet 67P was 3.6 au from the Sun and cometary activity was weak, until just before the comet reached its closest point from the Sun and experienced maximum activity in May 2015 at 1.7 au, as long as the VIRTIS-M infrared channel could record measurements (*6*). Analyses of the VIRTIS-M reflectance spectra ascribed the darkness and slope to a refractory polyaromatic carbonaceous component mixed with opaque minerals (anhydrous Fe-sulfides and Fe-Ni alloys), but the carrier of the 3.2-μm feature was unknown (*3*). Water ice contributes to this absorption on some parts of the surface (*6-9*), causing a broadening and deepening of the absorption feature from 2.7 to 3.1 μm, but cannot explain the entire feature (*3, 10*). Except in specific ice-rich areas, the surface of the cometary nucleus is uniform in composition, with a predominance of non-ice materials *(9)*. Semi-volatile materials of low molecular weight have been proposed as carriers of the 3.2-μm feature, with carboxylic (−COOH) bearing molecules or $NH_4^+$ ions being the most plausible candidates (*11*). However, a lack of reference spectral data for these compounds has prevented a firm attribution of the feature.

**Spectral identification of ammonium salts**

We conducted laboratory experiments to produce analogs of cometary surface material and measure their reflectance spectra under cometary-like conditions (low temperature and high vacuum) (*12*). Cometary dust is known to consist of aggregated sub-micrometer-sized (sub-μm) grains (*13*), and opaque iron sulfides are probable contributors to the low albedo of cometary nuclei (*11*, *14*). We therefore used sub-μm grains of pyrrhotite ($Fe_{1-x}S$, with $0 < x < 0.2$) mixed with different candidate compounds (carboxylic acid, ammonium salts) to test for the 3.2-μm feature. The pyrrhotite grains and candidate compounds were mixed in liquid water, then frozen to obtain ice-dust particles (*12*). By sublimating these particles in a thermal vacuum chamber, we formed very porous mixtures made of sub-μm grains (hereafter "sublimate residues", Figs. S1 and S2). This is representative of the process we expect at the surface of a cometary nucleus, and the resulting textures strongly influence the band depths of the reflectance spectra (*15*) which we attempt to reproduce to allow quantification of the components of the cometary surface (*12*).

Figure 1A shows the reflectance spectrum of the sublimate residue made of pyrrhotite grains mixed with ≲ 17 wt% (≲ 43 vol%) ammonium formate ($NH_4^+$ $HCOO^-$). Figure 1 also shows an average comet spectrum from a combination of VIRTIS-M observations taken between August to September 2014, when Rosetta was at a distance of 50 to 350 km from the nucleus and 67P was 3.6 to 3.3 au from the Sun (*12*). The position of the cometary absorption band, its asymmetric shape and the minima at 3.1 and 3.3 μm all match the absorption bands due to the N−H vibration modes of $NH_4^+$ in ammonium formate (Fig. 1A, Fig. S5). The spectral resolving power $RP \equiv \lambda/\Delta\lambda_{\mathrm{resolution}}$ (where $\lambda$ is the wavelength, and $\Delta\lambda_{\mathrm{resolution}}$ is the spectral resolution) at 3.5 μm is $RP = 90$ for the laboratory spectrum and 233 for the VIRTIS spectrum. The residual cometary spectrum, obtained by dividing the comet spectrum by the ammonium salt spectrum, is flat from 3.05 to 3.35 μm (Fig. 1A), but features from 3.35 to 3.60 μm indicate the additional presence of C−H stretching modes in carbonaceous compounds (*10*). The proximity of these C−H modes, the limited spectral resolution, and the limited spectral sampling impede a search for a weaker N−H mode of ammonium salts centered around 3.50 μm (Fig. S5, Table S2). Other differences between these spectra are due to the contribution of additional compounds (possibly traces of water ice around 3.0 μm) and/or differing properties of the salts present on the cometary surface (concentration, mixing, counter-ions etc.). Other candidate compounds we investigated – fine water ice grains, carboxylic acid, or hydrated minerals – do not match the 3.2-μm feature (Fig. 1B). Figure 2 shows the laboratory spectra of five ammonium salts we investigated. Ammonium formate, ammonium sulfate or ammonium citrate all reproduce the 3.1 and 3.3 μm absorption bands observed on the comet. For ammonium carbamate and ammonium chloride, the corresponding bands are shifted to longer wavelengths or have different spectral shapes (Fig. 2).

The similarity of band shapes and positions leads us to conclude that $NH_4^+$ in ammonium salts is the main responsible for the 3.2-μm feature. The counter-ion (here the anion) is not fully constrained. The identification of HCOOH by the Rosetta Orbiter Spectrometer for Ion and Neutral Analysis (ROSINA) mass spectrometer (*16*) and the spectrum of ammonium formate ($NH_4^+$ $HCOO^-$) in Figure 2 make it a favored candidate.

**Origins of ammonium salts**

There are several potential pathways for the synthesis of ammonium salts present at the surface of comet 67P. Ammonia ($NH_3$) has a high proton affinity allowing it to transform easily into ammonium ($NH_4^+$), either in the gas (*17*) or in the solid phase. Ammonium and potential counter-ions ($HCOO^-$, $CN^-$, $OCN^-$, etc.) may be produced by acid-base reactions of ammonia ($NH_3$) with the corresponding acids (HCOOH, HCN, HNCO, etc.) or by nucleophilic addition of $NH_3$ with $CO_2$ or $H_2CO$, even at cryogenic temperatures in the solid phase (*18, 19*). These reactions have low activation energies and do not require an external source of photons, electrons or cosmic rays (*18*). Some ions (e.g. $OCN^-$, $HCOO^-$) can be produced at 10-14 K, but most of the ions we consider (e.g. $NH_2COO^-$, $CN^-$) are produced at higher temperatures (*19-21*). Astronomical observations of interstellar ices have likely identified $OCN^-$ (*22-25*) and possibly detected $NH_4^+$ (*25-28*). Ammonium salts can be formed upon sublimation of water ice containing $NH_4^+$ and counter ions (*19, 29*). It is possible that the $NH_4^+$ detected on comet 67P could be inherited from interstellar ices. In that case, the ammonium salts would be produced during further thermal processing of the ices, either in the protoplanetary disk (*25*) or during the sublimation of the ices in the cometary nucleus, by a process similar to the one simulated in our laboratory experiments.

To our knowledge, the production of ammonium salts via a gas phase reaction under astrophysical conditions has never been reported in the literature. Solid-state reactions appear more likely, because proton transfer or nucleophilic addition are highly facilitated by a dust surface and a solvent such as ice (*30, 31*).

**Comparison with other small bodies**

The 3.2 µm absorption feature of comet 67P shares similarities with the 3-µm features observed on several asteroids, including the position and width of the band from 2.9 to 3.6 µm and the reflectance minimum at 3.1-3.2 µm (Fig. 3). However, the bands observed on most of these asteroids are distinct from the one of comet 67P, having a different shape and no secondary minimum at 3.3 µm (Fig. 3). Nevertheless, these spectra are compatible with the presence of ammonium salts, if the spectral differences are due to the environmental conditions at the surface of these small bodies (Fig. S7). The spectra of asteroids (24) Themis and (52) Europa are representative of objects found in the asteroid Main Belt (*1, 32*) and in orbit around Jupiter, e.g. the irregular moon Himalia (*33*). The dwarf planet (1) Ceres has ammonium-bearing minerals on its surface, with absorption features at 2.72 and 3.06 µm (Fig. 3), mostly in the form of phyllosilicates but with smaller amounts of salts (*34*). Our identification of ammoniated salts on a comet supports the hypothesis that materials on Ceres may have originated from the outer Solar System (*34*).

**Volatility of ammonium salts**

There is weak evidence for ammonium salts in meteorites, micrometeorites and interplanetary dust particles (IDPs) (*35*, *36*). Because these salts are more volatile than most refractory material, they might not be preserved during atmospheric entry of small particles and/or during long periods of time under terrestrial environmental conditions (*35*). Ammonium salts contained in grains ejected from cometary nuclei may react and/or sublimate when heated by the Sun and act as distributed sources of gases in comae, the envelopes of gas and dust around cometary nuclei (*37*). This could explain observed increases of $NH_3$ and HCN when some comets reach short heliocentric distances (< 1 au from the Sun), such as comet C/2012 S1 (ISON) which experienced multiple outbursts as it disrupted inside ~0.8 au (*38, 39*). On comet 67P, the decomposition of ammonium formate ($NH_4^+$ $HCOO^-$) could produce formamide ($NH_2CHO$), which has been detected by the ROSINA instrument and compatible with the mass spectrum measured by the Cometary Sampling and Composition (COSAC) experiment on Rosetta's Philae lander (*40-42*). The volatility of an ammonium salt strongly depends on the anion; e.g. at a pressure of 1 atm, ammonium formate decays at 389 K, ammonium sulfate at 553 K and ammonium chloride at 611 K (*43*). Under simulated astrophysical conditions ($10^{-8}$ mbar), the sublimation temperatures of ammonium salts are 160-180 K for ammonium cyanide ($NH_4^+$ $CN^-$) (*44*), 200-230 K for ammonium formate ($NH_4^+$ $HCOO^-$) (*30, 45*), and 230-260 K for ammonium carbamate ($NH_4^+$ $NH_2COO^-$) (*46*). Because of these differences of volatility, the composition of ammonium salts observed on comet or asteroid surfaces, and the gases produced by their decomposition, may change with the heliocentric distance.

**Nitrogen budget of comet 67P**

Rosetta's COmetary Secondary Ion Mass Analyzer (COSIMA) collected coma dust grains 10 to 30 km from comet 67P's nucleus and measured their composition (*47*). Ammoniated salts were not detected, possibly because any semi-volatile compounds present in the dust grains would have sublimated during the multiple-day-long pre-analysis storage of the particles at 283 K (*47*). If ammonium salts had been lost from the dust grains analyzed by COSIMA, their measured nitrogen-to-carbon ratio (N/C) would be a lower limit, missing the contribution of the semi-volatile nitrogen-bearing salts. COSIMA measured an average N/C of 0.035 ± 0.011, similar to the ratio found in the insoluble organic matter extracted from carbonaceous chondrite meteorites and in most micrometeorites and interplanetary dust particles (*47*), but significantly lower than the solar N/C value of 0.29 ± 0.12 (*48*). Similar depletions in nitrogen compared to the Sun have been found in the refractory dust and gas phases of other comets (*49–51*).

We propose that ammonium salts may constitute a substantial nitrogen reservoir in comet 67P, and possibly other comets and small bodies. The 3.2-µm band observed in the spectrum of comet 67P is 5 to 20% less deep than the band of ammonium formate in our sublimate residues (*12*) (Fig. S6). Assuming that the physical parameters controlling the light scattering (mixing modes, grain sizes etc.) of the sublimate residues are similar to that of the cometary surface, we derive an upper limit of the abundance of salts in the dark surface material of the comet of ~40 vol%. The dark surface material is a mixture of approximately 45 wt% organic (~1 g/cm$^3$) and 55 wt% mineral (~3.4 g/cm$^3$) components, estimated from COSIMA measurements (*52*). Taking into account this composition, we derive an upper limit of the mass fraction of ammonium salts mixed with the dust of about 40 wt%, but we cannot determine the surface abundance of ammonium salts on the comet exactly (*12*). If the mass fraction of ammonium formate ($NH_4^+$ $HCOO^-$) is 5 wt% in the cometary dust, the total atomic nitrogen in the comet is distributed as approximately 47% N in ammonium salts, 52% N in refractory organic matter and 1% N in volatiles (Fig. 4); the whole comet would then have a N/C ratio of about 0.06 (Fig. 5). If there is a mixture of several ammonium salts, then $NH_4^+$ would have a range of counter-ions, some of them N-bearing, which would raise the N/C ratio. Fig. 5 shows how the inferred N/C ratio of the comet increases with the assumed concentration of ammonium salts in the dust and depends on the nature of the counter-ions of $NH_4^+$.

**Implications**

The identification of ammonium salts on comet 67P shows that this comet, and possibly others, could have a N/C ratio higher than previously thought. If ammonium salts are a substantial repository of nitrogen, the assessment of their $^{14}N/^{15}N$ isotopic ratio and its comparison to the protosolar ratio could elucidate the incorporation and evolution of nitrogen in the early Solar System (*53*). If ammonium salts were also present in sufficient abundance in planetesimals during the early Solar System, they would have provided a solid form of nitrogen stable closer to the Sun than $N_2$ and $NH_3$ ices, and therefore available for planetary accretion (*54*). Abundant ammonium salts would have lowered the melting point of water ice in the subsurface of icy bodies (*55*). When mixed in liquid water, ammonium salts are known to have roles in potential prebiotic reactions, such as the formation of pyrimidine and purine nucleobases (*56*), the production of amino acids (*57*), the phosphorylation of nucleosides (*58*) or the formation of sugar molecules (*59*).

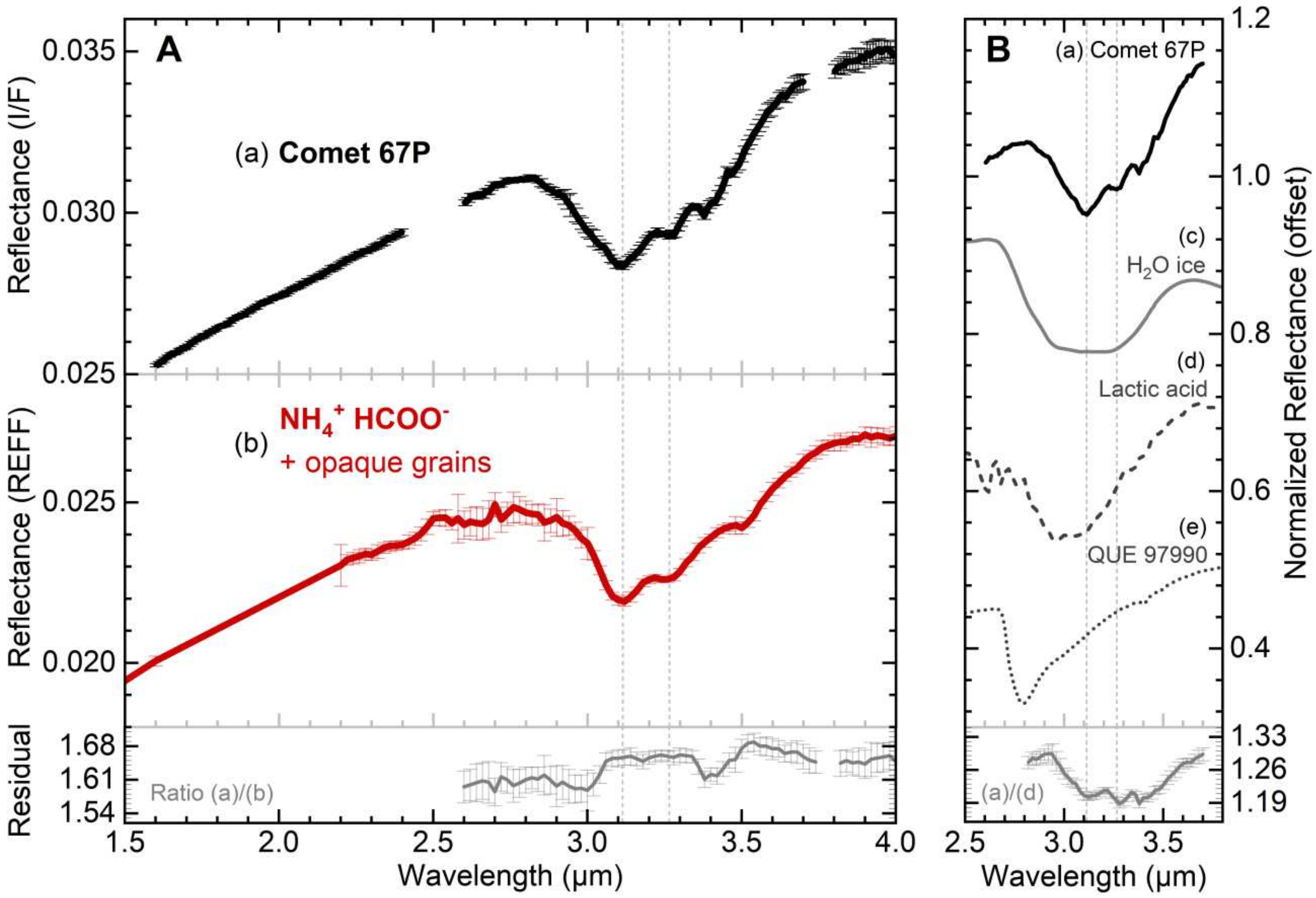

**Fig. 1. Comparison of $NH_4^+$ $HCOO^-$ spectrum with the average spectrum of comet 67P.** (**A**) The average reflectance spectrum of comet 67P in the 3.0 µm region ((**a**), black line, the vertical scale is given in radiance factor I/F) and the spectrum of a sublimate residue containing ≲ 17 wt% ammonium formate mixed with ≳ 83 wt% pyrrhotite grains at 170-200 K ((**b**), red line, the vertical scale is given in reflectance factor REFF). Fig. S5 shows the same spectra overlain on each other. Gaps in the comet 67P spectrum are due to the instrument's diffraction order sorting filters. Both spectra have the same shape and minima at 3.1 and 3.3 µm (dashed gray vertical lines). (**B**) Reflectance spectra (normalized at 2.5 µm) of other compounds that do not match the comet 67P spectrum: **(c)** a model spectrum of 1 µm diameter pure water ice grains (solid line) (Hapke model (*60*), using optical constants at 145 K (*61*), spectrum normalized, scaled by a factor of 0.14, and offset by -0.08). **(d)** a measured spectrum of a sublimate residue containing ≲ 17 wt% lactic acid mixed with pyrrhotite grains at 170-200 K (dashed line, offset by -0.35 *(12)*). **(e)** spectrum of the primitive carbonaceous chondrite meteorite QUE 97990, which is rich in hydrous silicates, measured under 400-475 K and high vacuum (dotted line, offset by -0.55 (*62*)). Between 2.5 and 2.8 µm, the spectra of sublimate residues are affected by measurement artifacts due to the presence of water vapor in the optical path. Residual spectra, calculated by dividing the comet 67P spectrum (a) by the experimental spectrum (b) or (d), are shown at the bottom of each panel. Error bars indicate the ±1σ uncertainty.

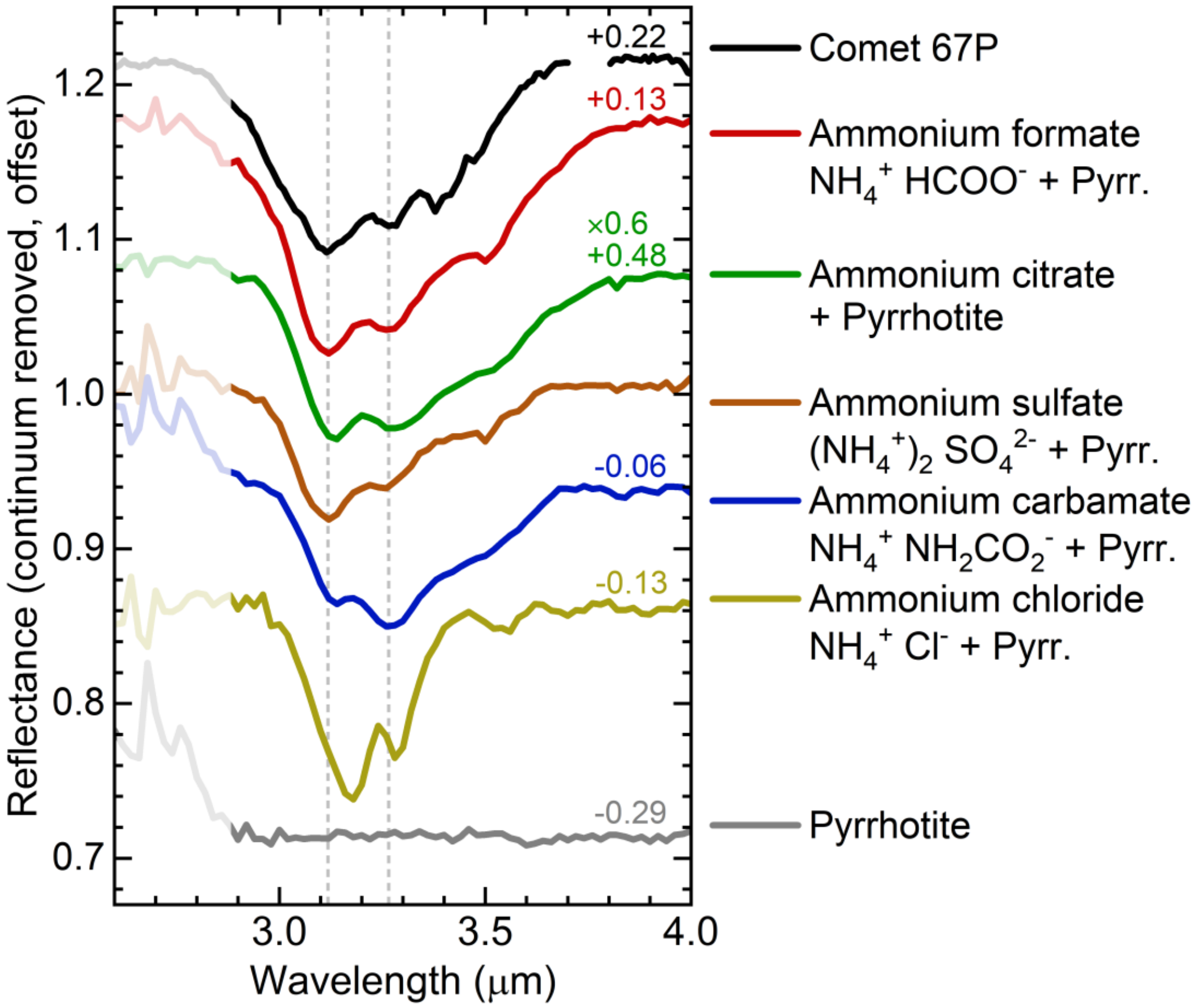

**Fig. 2. Reflectance spectra of several ammonium salts.** Continuum-subtracted reflectance spectra of sublimate residues made of ammonium salts (colored lines) mixed with pyrrhotite grains (gray line) measured in high vacuum at 170-200 K, compared to the observed average spectrum of comet 67P (black line). Ammonium formate, citrate and sulfate are the closest matches to the absorption features in comet 67P. We attribute the other absorption features in the comet spectrum at 3.35-3.6 µm to C−H stretching modes of organic compounds (*10*). Between 2.6 and 2.8 µm, the laboratory spectra are affected by measurement artifacts due to the presence of water vapor in the optical path. The mass fractions of the salts mixed with pyrrhotite are $\lesssim 9$ wt% for ammonium sulfate and chloride, $\lesssim 17$ wt% for ammonium formate and carbamate and $\lesssim 23$ wt% for ammonium citrate (the latter has been scaled by a factor of 0.6 for display). Figure S6 shows these spectra with uncertainties and prior to continuum subtraction.

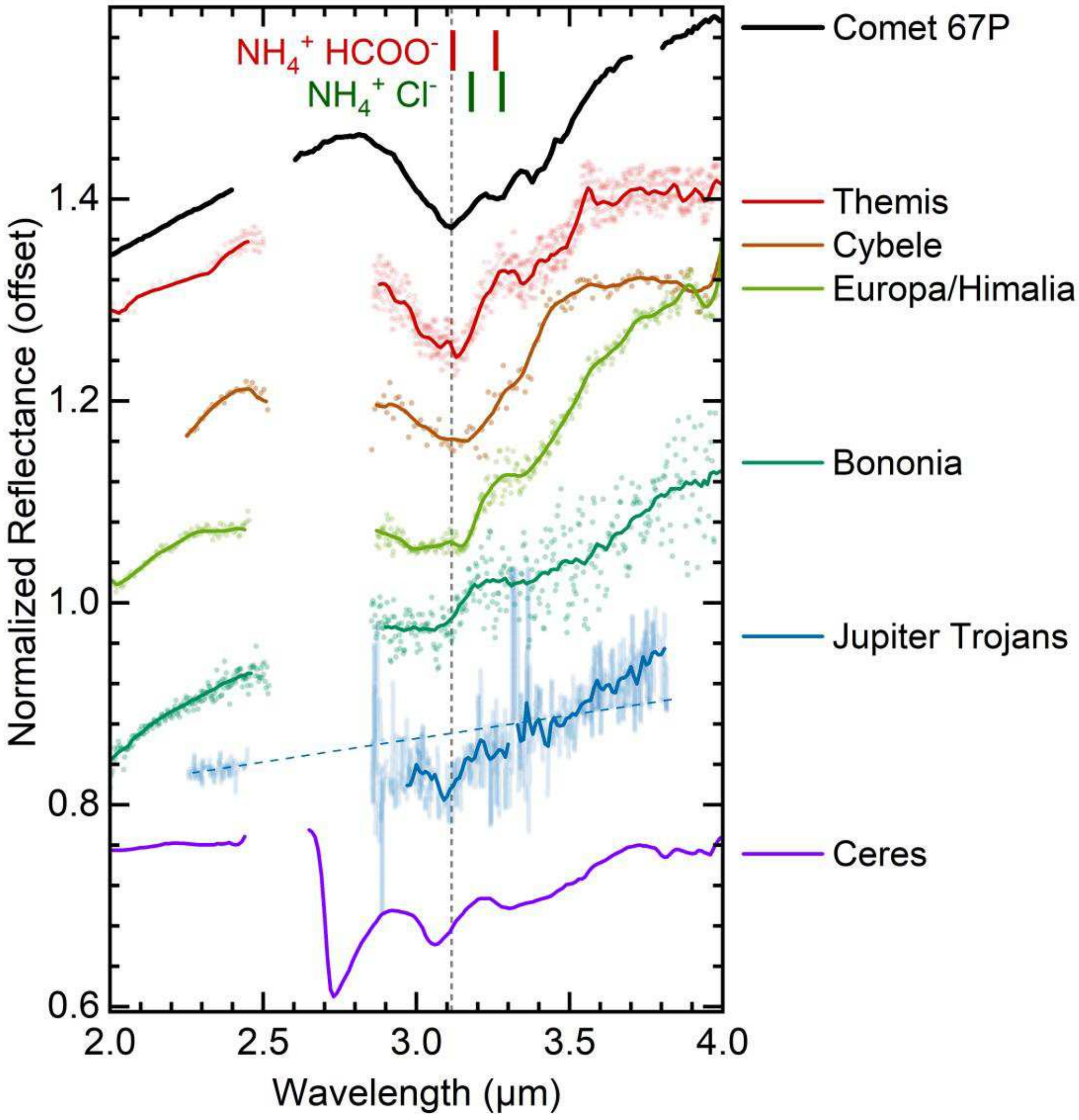

**Fig. 3. The spectrum of comet 67P compared to other Solar System bodies.** Reflectance spectra normalized at 2.9 µm are of comet 67P (offset by +0.45); the Main Belt asteroids (24) Themis (*63*) (offset by +0.31), (65) Cybele (*64*) (offset by +0.19), (52) Europa (*32*) and (361) Bononia (*32*) (offset by +0.07); the average spectrum of six Jupiter Trojan asteroids (*65*, their 'less red' group) (divided by 3, and offset by +0.49); and the average spectrum of (1) Ceres (*34*) (scaled by a factor of 0.5, and offset by +0.19). Jupiter's irregular moon Himalia has a spectrum, which is almost indistinguishable from (52) Europa (*33*). For each spectrum, the dots are the observational data (communicated by the authors for Europa and Bononia, or digitized for the others) and the solid lines are running average spectra. The blue dashed line shows the averaged extrapolation of the six Jupiter Trojan's K-band spectra (*66*). The gray dashed line shows the position of the band at 3.11 µm on comet 67P spectrum. The red and green vertical marks indicate the positions of the maxima of absorption of ammonium formate and ammonium chloride respectively, shown on Fig. 2. Absorption features around 3.1-3.2 µm on some of these bodies are similar to the ammonium salt features on comet 67P. Ceres exhibits different features, which are due to ammoniated phyllosilicates *(34)*.

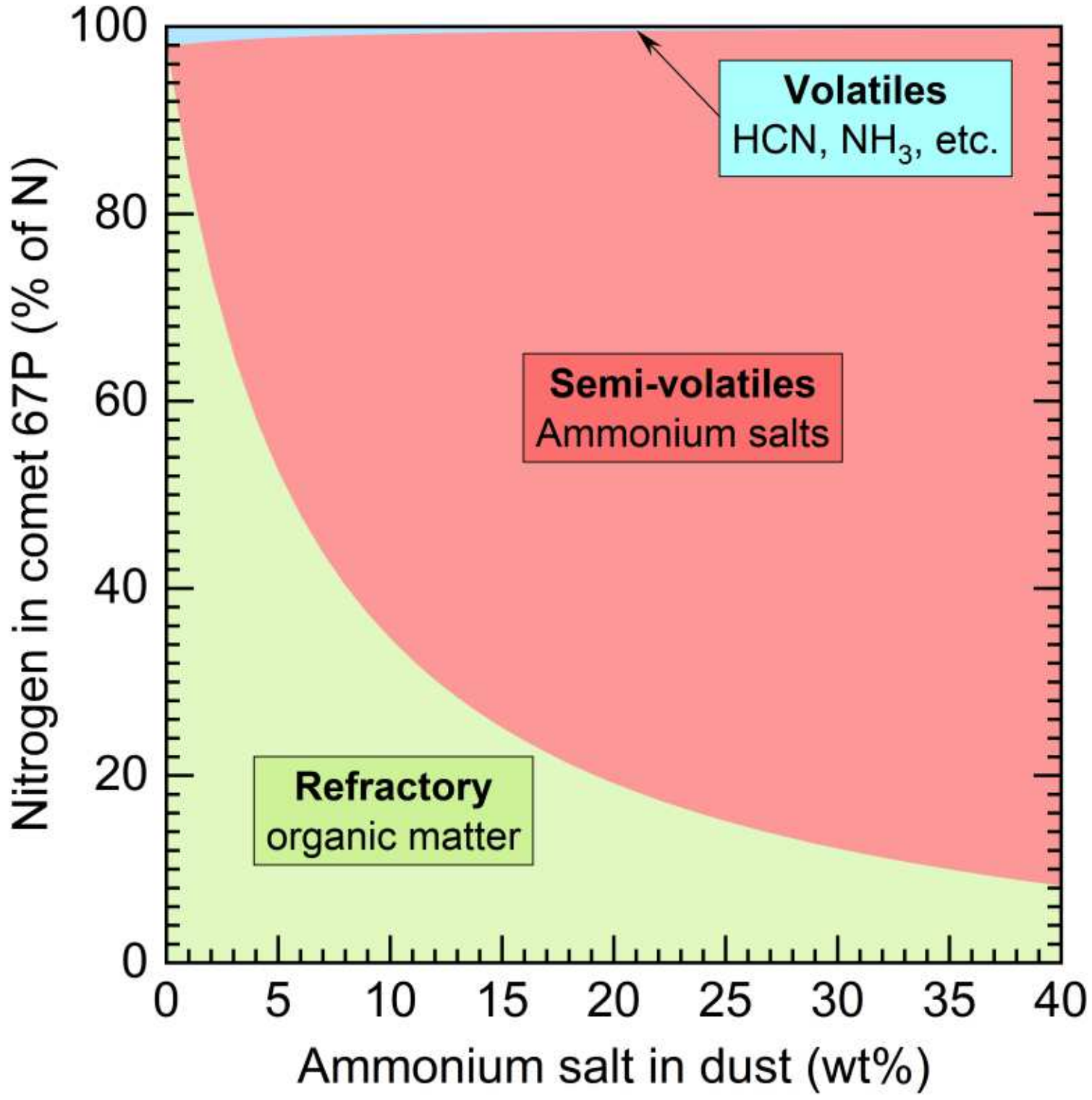

**Fig. 4. Distribution of nitrogen in comet 67P.** Nitrogen in comet 67P is found in the refractory organic matter, in semi-volatile ammonium salts and in volatile molecules. This diagram shows an estimation of how all the nitrogen atoms in comet 67P are distributed among these three reservoirs, depending on the mass fraction of ammonium salt in the dust (composed of minerals, refractory organic matter, and salts). These calculations are based on observations from several Rosetta instruments (*12*). If the cometary dust contains more than few percent of ammonium salts, then they form a substantial reservoir of nitrogen in comet 67P.

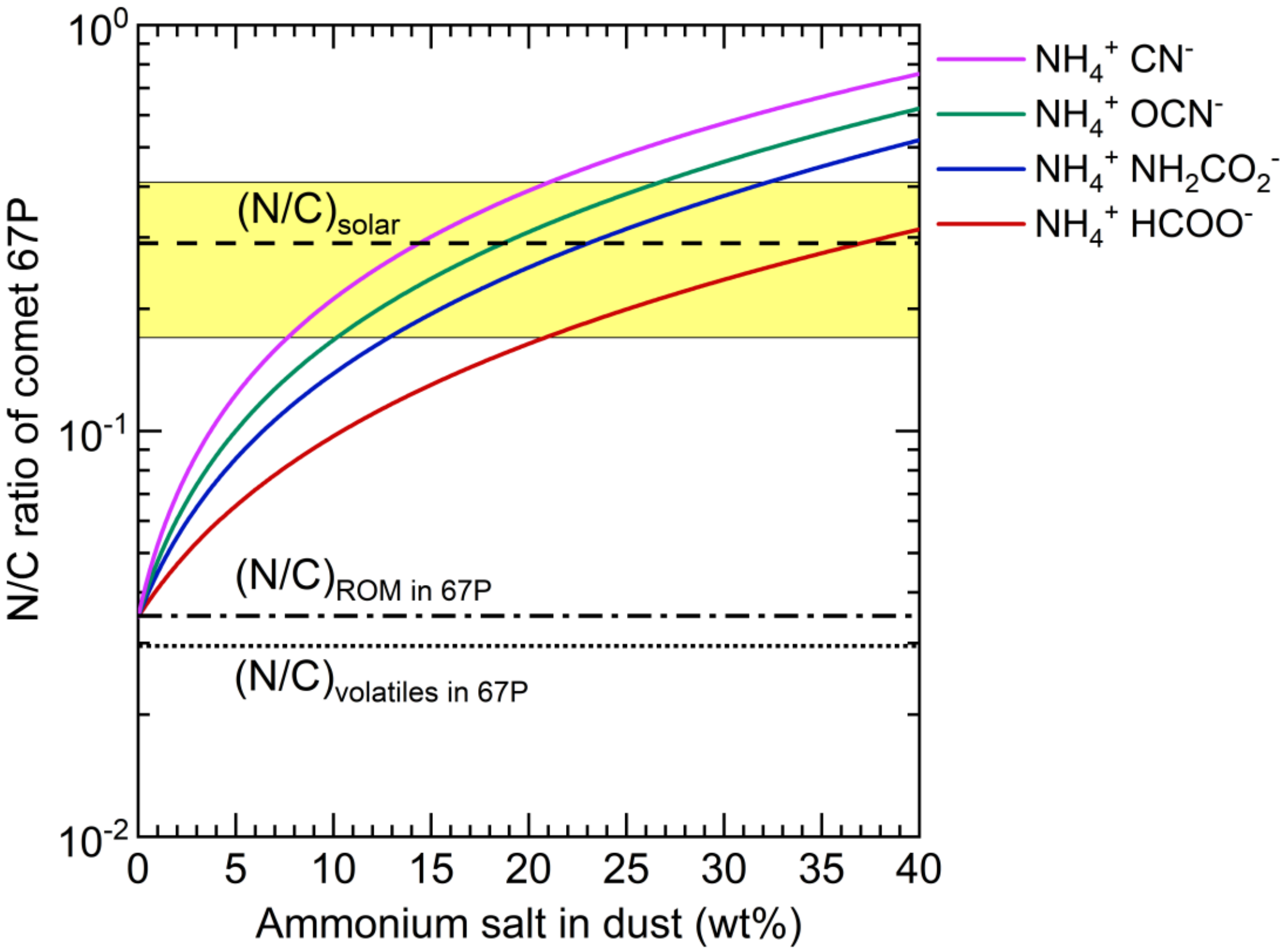

**Fig. 5. Nitrogen-to-carbon ratio in comet 67P.** The N/C ratio in comet 67P compared to the solar value (dashed line and yellow area indicating the ±1σ uncertainty (*48*)). The colored lines show the contributions of nitrogen in plausible ammonium salts as a function of the mixing with dust (*12*) which are added to the nitrogen in the refractory organic matter (ROM) (*47*) and in volatile species (*16*) (dashed-dotted and dotted lines respectively). Depending on the counter-ion, the presence of 10 to 30 wt% of ammonium salts in the dust would raise the N/C ratio of comet 67P enough to be consistent with the solar value.

**Acknowledgement:** O.P., I.I. and E.Q. thank Frédéric Charlot of the Consortium des Moyens Technologiques Communs (CMTC) at the Institut National Polytechnique de Grenoble (INP) for the Scanning Electron Microscopy images of the samples. We acknowledge Mathilde Faure for work on the VIRTIS data, and the collaboration of the International Space Science Institute (ISSI) international team number 397 "Comet 67P/Churyumov-Gerasimenko Surface Composition as a Playground for Radiative Transfer Modeling and Laboratory Measurements". O.P. thanks Cédric Pilorget for insightful comments. **Funding:** O.P. acknowledges a post-doctoral fellowship from the Centre National d'Etudes Spatiales (CNES). I.I. acknowledges a thesis grant from the Lembaga Pengelola Dana Pendidikan (LPDP) Indonesian scholarship. L.M. acknowledges the Deutsche Forschungsgemeinschaft (DFG) grant MO 3007/1-1. D.K. acknowledges DFG-grant KA 3757/2-1. The work of O.P., I.I., E.Q., P.B., B.S. and L.B. was supported by the CNES and the French Agence Nationale de la Recherche (programme Classy, ANR-17-CE31-0004). P.B. acknowledges funding from the European Research Council under the SOLARYS grant (ERC-CoG2017-771691). P.H.B. acknowledges the LabEx Observatoire des Sciences de l'Univers de Grenoble OSUG @ 2020 for funding. S.P. is supported by Université Grenoble Alpes (UGA) Initiatives de Recherche Stratégiques (IRS) and Initiatives d'Excellence UGA (IDEX UGA). The development of the CarboN-IR environmental chamber was supported by the French national program of planetology (PNP), and the development of the reflectance goniometers was supported by the University of Grenoble Alpes (Initiative de Recherche Stratégique). Université Grenoble Alpes (UGA) and CNES supported the instrumental facilities and activities at IPAG. The following institutions and agencies also supported the Rosetta mission: the Italian Space Agency (ASI - Italy), Centre National d'Etudes Spatiales (CNES - France), Deutsches Zentrum für Luft- und Raumfahrt (DLR - Germany), the National Aeronautic and Space Administration (NASA - USA) Rosetta Program and the Science and Technology Facilities Council (UK). VIRTIS was built by a consortium including Italy, France and Germany, under the scientific responsibility of the Istituto di Astrofisica e Planetologia Spaziali of INAF, Italy, which also guided scientific operations. The development of the VIRTIS instrument was funded and managed by ASI, with contributions from Observatoire de Meudon, financed by CNES, and from DLR. A.R., M.C., G.F., F.C., M.C.D.S., A.L., E.P. and F.T. acknowledge financial support from the National Institute for Astrophysics (INAF, Italy) and the Italian Space Agency (ASI, Italy) through contract number I/024/12/2. Computational resources were provided by INAF-IAPS through the DataWell project. The work of A.P. and N.T. has been carried out within the framework of the National Centre of Competence in Research (NCCR) PlanetS supported by the Swiss National Science Foundation. **Author Contributions:** O.P. and I.I. carried out the laboratory experiments. O.P. wrote the manuscript with assistance from P.B., E.Q., and B.S., who also contributed to interpreting the results. E.Q., P.H.B., A.F. and P.B. calculated the nitrogen distribution and N/C ratio. O.P., E.Q. and A.F. produced the figures. A.R., M.C. and G.F. provided the calibrated average reflectance spectrum of comet 67P and wrote parts of the supplementary material. All co-authors contributed to the preparation of the manuscript. O.B., S.P., B.R. and L.F. contributed to the experimental work. N.T. and A.P. contributed to the early development of spectroscopic studies of sublimate residues.

**Competing interests:** We declare no competing interests. **Data and materials availability:** The laboratory reflectance spectra are available online in the Grenoble Astrophysics and Planetology Solid Spectroscopy and Thermodynamics (GhoSST) database (*67, 68*). The average reflectance spectrum of comet 67P measured by VIRTIS is available at https://www.nature.com/articles/s41550-019-0992-8#Sec10 (*10*).

**Supplementary Materials:**

Materials and Methods

Figures S1-S9

Tables S1-S2

References (*69-89*)

# Supplementary Materials for

## Ammonium salts are a reservoir of nitrogen on a cometary nucleus and possibly on some asteroids

Olivier Poch*, Istiqomah Istiqomah, Eric Quirico, Pierre Beck, Bernard Schmitt, Patrice Theulé, Alexandre Faure, Pierre Hily-Blant, Lydie Bonal, Andrea Raponi, Mauro Ciarniello, Batiste Rousseau, Sandra Potin, Olivier Brissaud, Laurène Flandinet, Gianrico Filacchione, Antoine Pommerol, Nicolas Thomas, David Kappel, Vito Mennella, Lyuba Moroz, Vassilissa Vinogradoff, Gabriele Arnold, Stéphane Erard, Dominique Bockelée-Morvan, Cédric Leyrat, Fabrizio Capaccioni, Maria Cristina De Sanctis, Andrea Longobardo, Francesca Mancarella, Ernesto Palomba, Federico Tosi

*Correspondence to: olivier.poch@univ-grenoble-alpes.fr

**This PDF file includes:**

## Materials and Methods

### Preparation of the cometary analogs

We performed experiments to measure the reflectance spectra of ammonium salts and a carboxylic acid when they are mixed in a porous matrix of opaque sub-micrometer-sized grains, designed to simulate the mixing mode found on a cometary nucleus. Previous work has shown that the sublimation of water ice particles containing inclusions of non-volatile components produces porous sublimate residues (*15*, *69*). We prepared sublimate residues of an opaque mineral and salts following the same protocol (*15*, *69*).

For the dust component, we used pure pyrrhotite ($Fe_{1-x}S$, with $0 < x < 0.2$) because it is an opaque mineral proposed as a possible darkening agent of the cometary nucleus at infrared wavelengths (*11*, *14*). Millimeter-sized pyrrhotite grains were purchased from Alfa-Aesar (ref. 42652) and ground to sub-micrometer sized grains (*14*, their section 2.2). Most of the ammonium salts and the carboxylic acid were purchased from Sigma-Aldrich: ammonium formate ≥ 99.995 % (ref. 516961), ammonium sulfate ≥ 99 % (ref. A4418), ammonium carbamate 99 % (ref. 292834). The ammonium chloride ≥ 99.5 % was from Carlo Erba (ref. 419417), L-lactic acid anhydrous 98 % from Alfa-Aesar (ref. L13242), and ammonium citrate dibasic ≥ 99 % from Fluka (ref. 09831) whose detailed formula is $(NH_4^+)_2\ CH_2COOH\text{-}C(OH)COO^-\text{-}CH_2COO^-$.

The salts or the carboxylic acid were dissolved in ultra-pure liquid water. For example, in the case of the ammonium formate, 0.04 g of salt powder was dissolved in 20 mL of ultra-pure water, resulting in a mass ratio of 0.002. Then, 0.2 g of sub-micrometer sized pyrrhotite grains were dispersed in this solution (1 wt%) via ultrasonication, to destroy the aggregates and obtain a homogeneous mixture (*69*, their section 2.2.2). The ice particles were prepared from this liquid mixture using the Setup for the Production of Icy Planetary Analogs – B (SPIPA-B) (*15*, their section 2.3.4.2). The SPIPA-B ice particles are spherical, with a mean diameter of 67 ± 31 µm as measured by Cryo-Scanning Electron Microscopy (*15*). The pyrrhotite grains and the salt are contained inside each of these particles, this mixing mode being referred to as "intra-mixture" (*15*). The produced ice/dust particles were placed in a cylindrical aluminum sample holder of 48 mm diameter and 5 mm deep, by direct sieving through a 400 µm sieve to exclude large agglomerates and to obtain homogeneous surface and internal density (~0.5 g/cm$^3$) (Fig. S1B).

The sample holder containing the ice/dust mixture was maintained at a temperature lower than 170 K and under a dry atmosphere in an insulating box containing liquid nitrogen before being placed inside the CarboN-IR simulation chamber (*70*). Inside the chamber, the sample was kept at 170-200 K using a helium cryostat and under high vacuum (< $10^{-5}$ mbar). The surface of the sample absorbs thermal infrared radiation emitted from the top and the walls of the chamber, so the water ice sublimates progressively. As the surface dehydrates, it becomes darker because of the formation of a sublimate residue composed of pyrrhotite grains and salts or acid (Fig. S1). After about 48 hours, all the water ice has sublimated (Fig. S1C). Scanning Electron Microscopy images of these sublimate residues show a porous network of sub-µm pyrrhotite grains where the salts or acid may be present as coating or cement between the grains (Fig. S2). For initial ice particles containing 1 wt% pyrrhotite and 0.2 wt% salt, we estimate that the sublimate residue is made of 83 wt% pyrrhotite and 17 wt% salt, but a fraction of the salt might have sublimated with the water so the quantity of salt in the final residue may be lower.

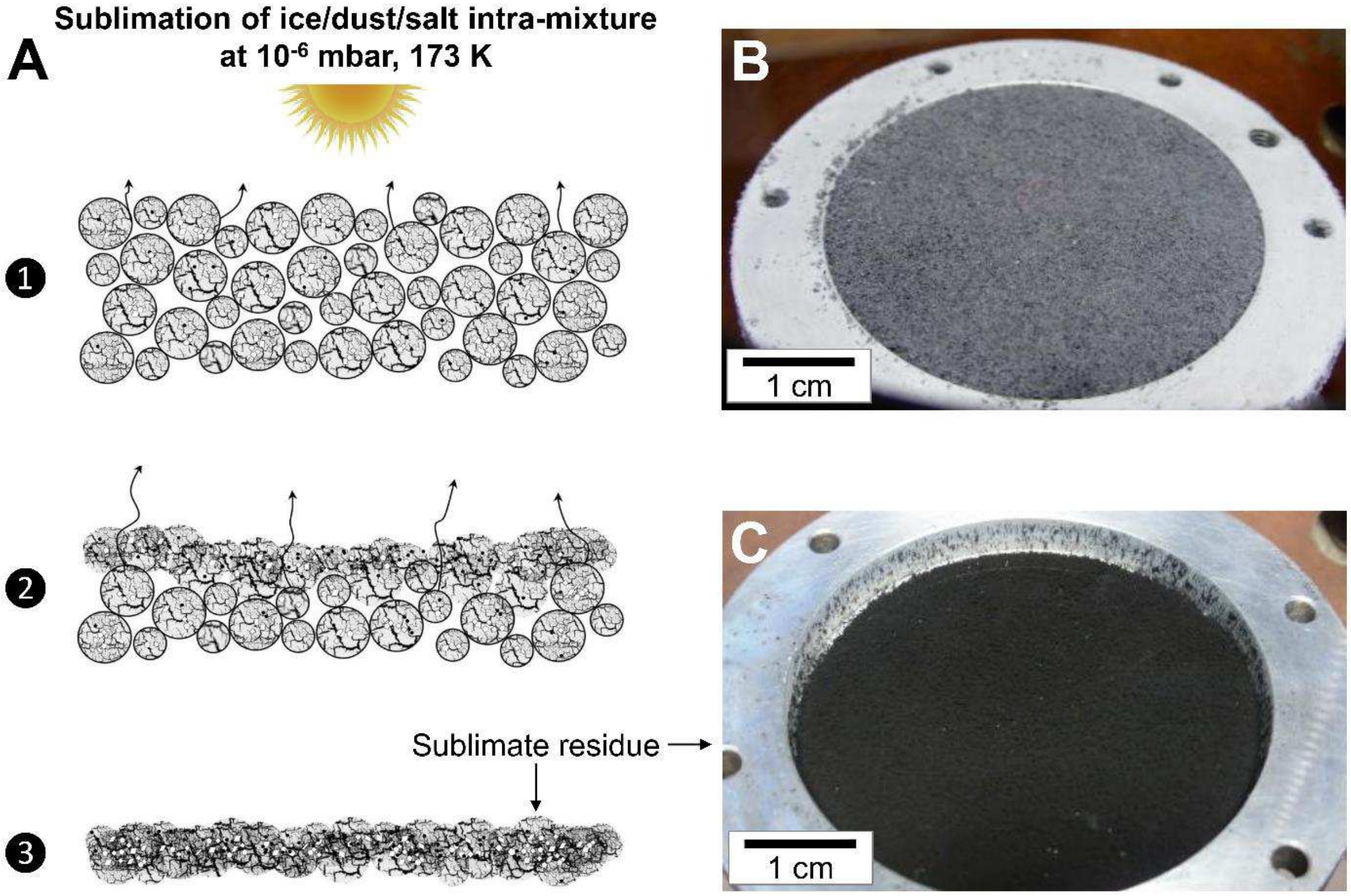

**Fig. S1. Production of a sublimate residue.** (**A**) Preparation of the cometary surface analog – a sublimate residue made of dust and salt − in the laboratory. Spherical water ice particles containing 1 wt% pyrrhotite and 0.2 wt% ammonium formate ($NH_4^+$ $HCOO^-$) are placed in a thermal vacuum chamber under high vacuum (< $10^{-5}$ mbar) and 170-200 K. The sublimation of the water, triggered by thermal infrared or visible light, results in the formation of a porous residue made of pyrrhotite grains containing 17 wt% or less of $NH_4^+$ $HCOO^-$. (**B**) Photograph of the initial sample made of water ice particles containing 1 wt% pyrrhotite and 0.2 wt% ammonium formate, before sublimation. (**C**) Photograph of the same sample after sublimation of the water ice.

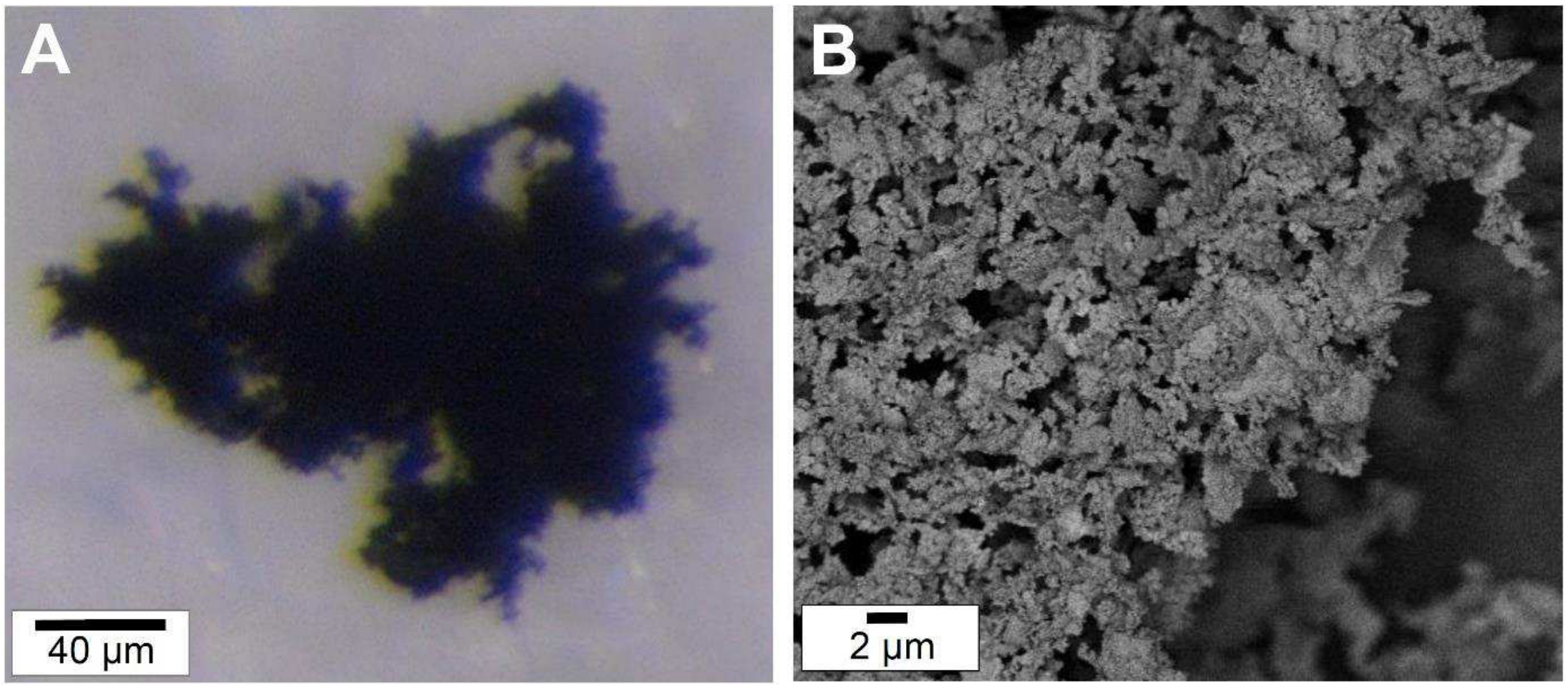

**Fig. S2. Micro-structure of a sublimate residue.** (**A**) Optical microscopy image of a fragment of the sublimate residue shown in Fig. S1C, showing its fluffy structure. (**B**) Backscattered electron image obtained by Scanning Electron Microscopy of the sublimate residue, showing sub-micrometric-sized grains of pyrrhotite arranged in a porous structure as a result of the freezing and sublimation processes. The ammonium salt probably coats and/or cements the pyrrhotite grains.

Reflectance measurements in the laboratory

The reflectance measurements were performed using the SpectropHotometer with variable INcidence and Emergence (SHINE) of the Institut d'Astrophysique et de Planétologie de Grenoble (IPAG). The instrument (*71*) consists of two goniometer arms: the first illuminates the sample with a monochromatic beam, provided by a monochromator, and the second holds two detectors covering the 0.5-5.0 µm spectral range. A transparent sapphire window allows the sample to be illuminated and viewed inside the CarboN-IR simulation chamber, positioned at the basis of the arms. SHINE was designed to measure bright samples of ice, but the sublimate residues rich in pyrrhotite are much darker samples. Therefore, we focused the illumination of the sample onto a smaller area of 6 mm in diameter, in order to increase the signal-to-noise ratio of our measurements (*72*, their section D.2). For the computation of the reflectance factor (REFF) (*60*) we also measured two surfaces made of Spectralon® and Infragold® (Labsphere Inc.) as references, corrected for their non-ideal spectral and photometric behavior (*73*). All the reflectance spectra were measured with a nadir incidence (illumination) and an emission (observation) angle of 30°. Spectral sampling $\Delta\lambda_{sampling}$ = 0.020 µm was constant throughout the overall wavelength range, but the spectral resolution varies, with higher resolutions at shorter wavelengths and lower resolutions at longer wavelengths. The spectral resolution was $\Delta\lambda_{resolution}$ = 0.019 µm in the wavelength range of 1.4-3.0 µm, and $\Delta\lambda_{resolution}$ = 0.039 µm in the range of 3.0-4.2 µm. The abrupt change of spectral resolution at 3.0 µm is due to a change of grating in the monochromator. The average resolving power over these wavelength intervals is thus $RP \equiv \lambda/\Delta\lambda_{resolution} \approx 100$ where $\lambda$ is the wavelength, which is lower than that of VIRTIS-M ($\lambda/\Delta\lambda_{resolution} \approx 230$). The absolute radiometric precision of the laboratory measurement was better than 3% over the whole spectral range. The spectrum shown in Figs. 1 and S5 is the average of 30 individual spectra measured over a period of 16 hours.

Reflectance spectrum of the average surface of comet 67P

The reflectance spectrum of comet 67P shown in Figs. 1, 2, S5, S6 and S7 was obtained from a recalibration of data acquired by the VIRTIS-M (Visual Infrared and Thermal Imaging Spectrometer – Mapper) Infrared (IR) channel (*2*), reducing instrumental artifacts and improving the radiometric accuracy. We summarize the data reduction below; full details have been published elsewhere *(10)*.

Average spectrum of comet 67P surface

The VIRTIS-M IR channel acquired hyperspectral images in the range 1 - 5 µm, across 432 spectral bands, and through 256 spatial samples of 250 µrad Instantaneous Field Of View (IFOV).

To calculate an average spectrum of the surface of comet 67P we consider the whole dataset acquired during the first mapping phase of the Rosetta mission in August-September 2014 (~2.7 million spectra) at a distance ranging from 350 to 10 km from the center of the nucleus. These observations revealed a spectrally uniform surface (*3, 4, 9*). Spatial pixels corresponding to shadowed areas are filtered out and not used in the processing because they contain a very low signal-to-noise ratio.

By deriving an average spectrum from the full dataset, we reduce the Poisson noise to a level negligible. However, systematic instrumental effects in the conversion from digital number to radiance, varying from sample to sample (linearly dependent on the signal) contaminate the spectra. They generate artifacts on the signal visible at small spectral scale when comparing spectra of comet 67P and the spectra of a different target observed by VIRTIS with the same spatial sampling (Fig. S3).

To remove such artifacts, we produced average spectra of the comet nucleus independently for each spatial sample. The same average has been computed for asteroid Lutetia, which Rosetta flew past during the cruise phase of the mission (*74*). The ratio between the average spectra of comet 67P and Lutetia has been calculated independently for each sample, to eliminate spectral artifacts while keeping information of the real features. We assume absorption bands in the comet 67P spectrum to lie in spectral ranges where the asteroid Lutetia spectrum is featureless (as shown by VIRTIS-H observations of Lutetia (*74*) and confirmed by applying the same process to a reference Mars spectrum available at the Planetary Science Archive https://archives.esac.esa.int/psa/#!Table%20View under the observation ID I1_00130974021). To obtain the absolute spectrum of comet 67P, the ratio of the 67P and the Lutetia spectra is then multiplied by a 9-degree polynomial model fitted to the Lutetia spectrum (which we label 'interp'), to produce an artifact-removed (AR) average spectrum of comet 67P for each sample (*s*) (Fig. S3):

$$\frac{I}{F}(\lambda,s)^{\mathrm{AR}}_{\mathrm{67P}} = \frac{\frac{I}{F}(\lambda,s)_{\mathrm{67P}}}{\frac{I}{F}(\lambda,s)_{\mathrm{Lutetia}}} \cdot \frac{I}{F}(\lambda,s)_{\substack{\mathrm{interp}\\ \mathrm{Lutetia}}} \quad \text{(S1)}$$

where *I/F* is the radiance factor (π times bidirectional reflectance), $\frac{I}{F}(\lambda,s)^{\mathrm{AR}}_{\mathrm{67P}}$ the radiance factor of the artifact-removed average spectrum of comet 67P, $\frac{I}{F}(\lambda,s)_{\mathrm{67P}}$ and $\frac{I}{F}(\lambda,s)_{\mathrm{Lutetia}}$ are the radiance factor of the absolute spectrum of comet 67P and Lutetia respectively, and $\frac{I}{F}(\lambda, s)_{\substack{\text{interp}\\ \text{Lutetia}}}$ is the radiance factor of the polynomial fit of the Lutetia spectrum.

The resulting average spectra are affected by a further source of non-Poissonian and non-systematic noise: due to the detector's architecture, the even and odd spectral bands are controlled by two independent sets of electronics. Small differences in their responses introduce an oscillation along the wavelengths (*65*). The response of the even bands is spuriously affected by the instrument temperature. Thus, we replaced the signals of the even bands by an average of the contiguous odd spectral bands. Nominally, the VIRTIS-M IR channel has a bandpass of 0.015 µm with a sampling of 0.010 µm. After removal of the even bands, the sampling is degraded to 0.020 µm.

Finally, the average spectrum of the comet 67P surface $\frac{I}{F}(\lambda)_{67\text{P}}^{\text{AR}}$ has been calculated by applying a median filter to $\frac{I}{F}(\lambda, s)_{67\text{P}}^{\text{AR}}$ over all samples.

The *I/F* uncertainty associated with each spectral band is obtained from the standard deviation over all $\frac{I}{F}(\lambda, s)_{67\text{P}}^{\text{AR}}$, as each sample can be considered as an independent detector. To exclude from this calculation signal variations due to different viewing geometry and illumination conditions of the comet nucleus as seen along the spectral slit, we calculated the standard deviation after scaling $\frac{I}{F}(\lambda, s)_{67\text{P}}^{\text{AR}}$ to the median spectrum $\frac{I}{F}(\lambda)_{67\text{P}}^{\text{AR}}$ using the ratio of the medians over the spectral bands as scaling factor. These uncertainties are assumed to represent the relative errors on the spectral shape.

Absolute calibration with star observations

Both VIRTIS-Rosetta and the Visual and Infrared Mapping Spectrometer (VIMS) on the Cassini-Huygens spacecraft measured the Arcturus star flux. Comparing two observations of Arcturus by VIRTIS-M-IR with six observations by VIMS (*76*) we performed an inter-calibration. The ratio of the average fluxes observed by the two instruments provides a correction factor as a function of wavelength, which can be applied to the final average spectrum of comet 67P (Fig. S3). We rely on VIMS consolidated calibration (*77*) because it was tested and improved over the duration of the Cassini mission.

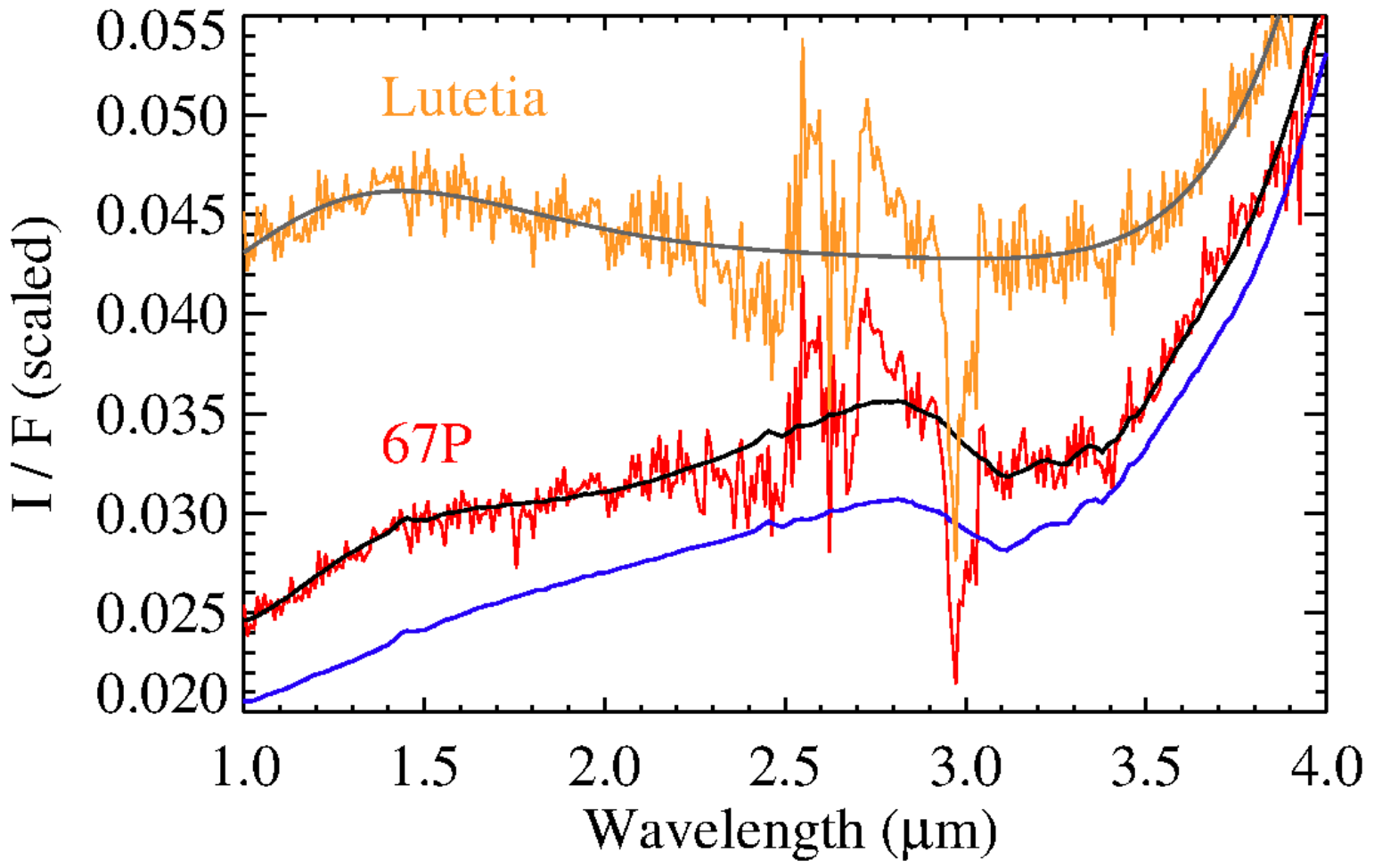

**Fig. S3. Absolute calibration of VIRTIS-M observations of comet 67P.** Average spectrum over all spatial pixels (~10000) that show illuminated areas of 67P nucleus acquired by a selected single spatial sample (red line). For comparison, the average spectrum of Lutetia for the same spatial sample, scaled for display (orange line) (original spectrum scaled by a factor of 0.56), and a polynomial model fitted to the data (gray line) are shown. The detector temperature and integration time for the observation of 67P and Lutetia are indicated in Table S1. The black line is the calibrated spectrum after artifact and odd-even effect removal, computed from the orange, red and gray curves using Equation S1. The final spectrum after application of the stellar calibration factor is represented as a blue line. After *(10*, their Supplementary Figure S3*)*, used with permission of the authors.

| Object | Detector DN range (3.0-3.5 µm) | Detector temperature (K) | Detector exposure time (s) | File names |
|---|---|---|---|---|
| **Asteroid Lutetia** | 7700 ± 300 | 87.74 | 0.7 | I1_00237396952 |
| **Comet 67P** | 8200 ± 300 | 87.69 ± 0.22 | 1, 1.5, 2, 3, 10 | (*10*) |

**Table S1. Parameters of VIRTIS-M detector.** Ranges of Data Numbers (DN) (in the spectral range 3.0-3.5 µm), detector temperature and exposure time during the observations of asteroid Lutetia and comet 67P by VIRTIS-M. The file name from the Planetary Science Archive (https://archives.esac.esa.int/psa/) is given as reference for the unique cube of Lutetia observation. For comet 67P, the data used to obtain the average spectrum is from more than 257 cubes measured from August to September 2014 (*10*).

Thermal emission removal

Thermal emission of the comet nucleus produces a steep spectral slope increasing longward of 3 µm. To compare the observed average spectrum with laboratory measurements, the contribution of the thermal emission must be estimated and removed. We modeled the total signal longward of 2.2 µm as the sum of the thermal emission and the solar flux reflected from the surface.

The thermal emission is modeled as gray body radiation, assuming no spectral variation of the directional emissivity (*60*). This assumption is consistent with the low albedo of the comet and its small variability in the spectral range considered (*78*). Temperature and effective emissivity are free parameters for the thermal emission modeling. The reflectance is modeled as a linear function with the corresponding intercept and spectral slope as free parameters.

The free parameters of the model are retrieved by a fitting procedure in the 2.2 – 2.8 µm and 3.6 – 4.15 µm spectral ranges, thereby excluding the wavelengths covering the broad 3.2 µm absorption. The modeled thermal emission has been removed over the entire spectral range longward of 2.2 µm (Fig. S4).

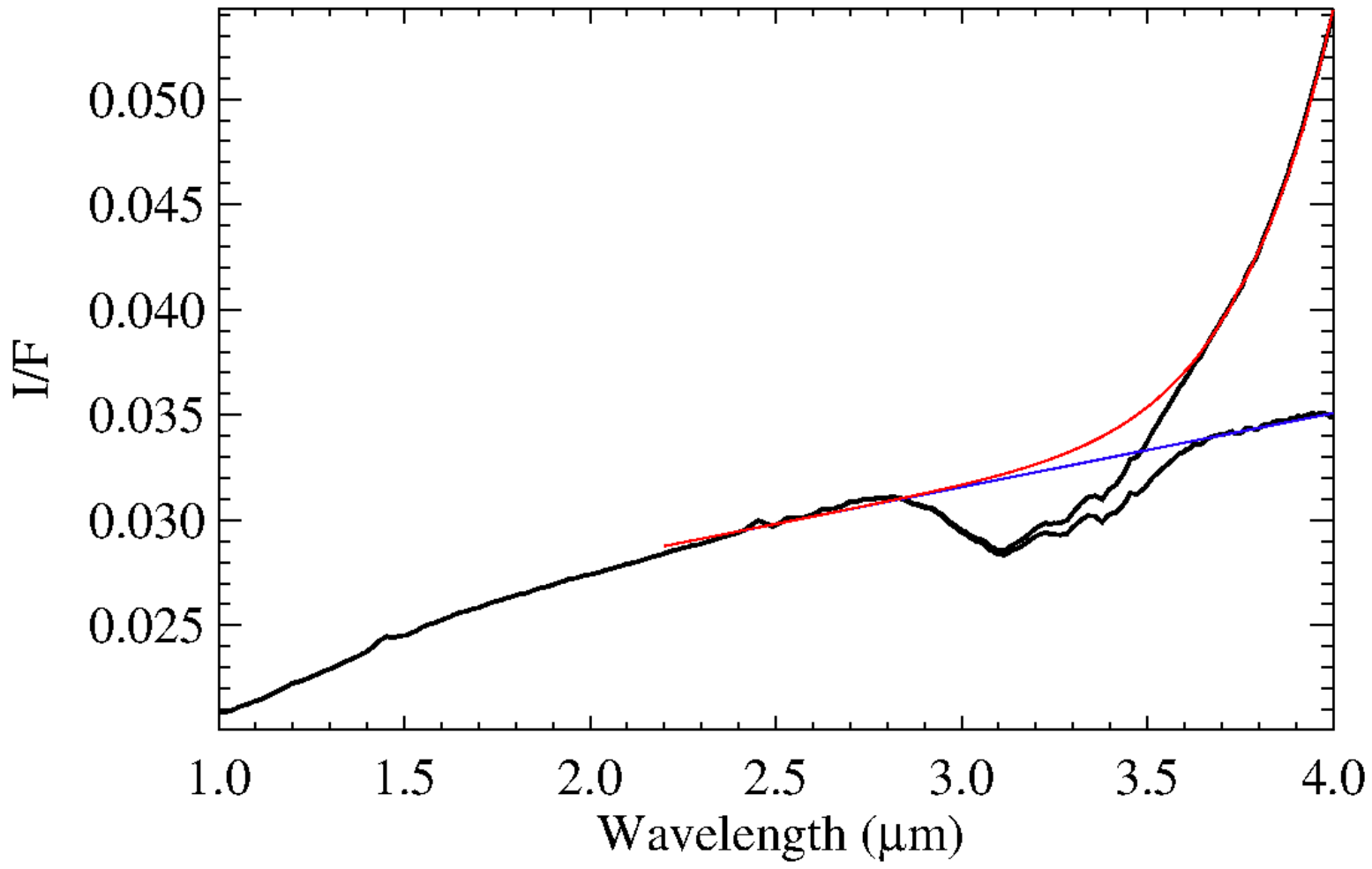

**Fig. S4. Thermal emission removal on VIRTIS-M observations of comet 67P.** The total modeled continuum signal (red line) is given by the sum of the modeled reflectance continuum (blue line) and the thermal emission. The latter is subtracted from the measured calibrated signal (upper black line), yielding the thermal emission removed spectrum (lower black line). Reproduced from *(10*, their Supplementary Figure S4*)*, used with permission of the authors.

Comparisons of experimental reflectance spectra with the average comet 67P spectrum

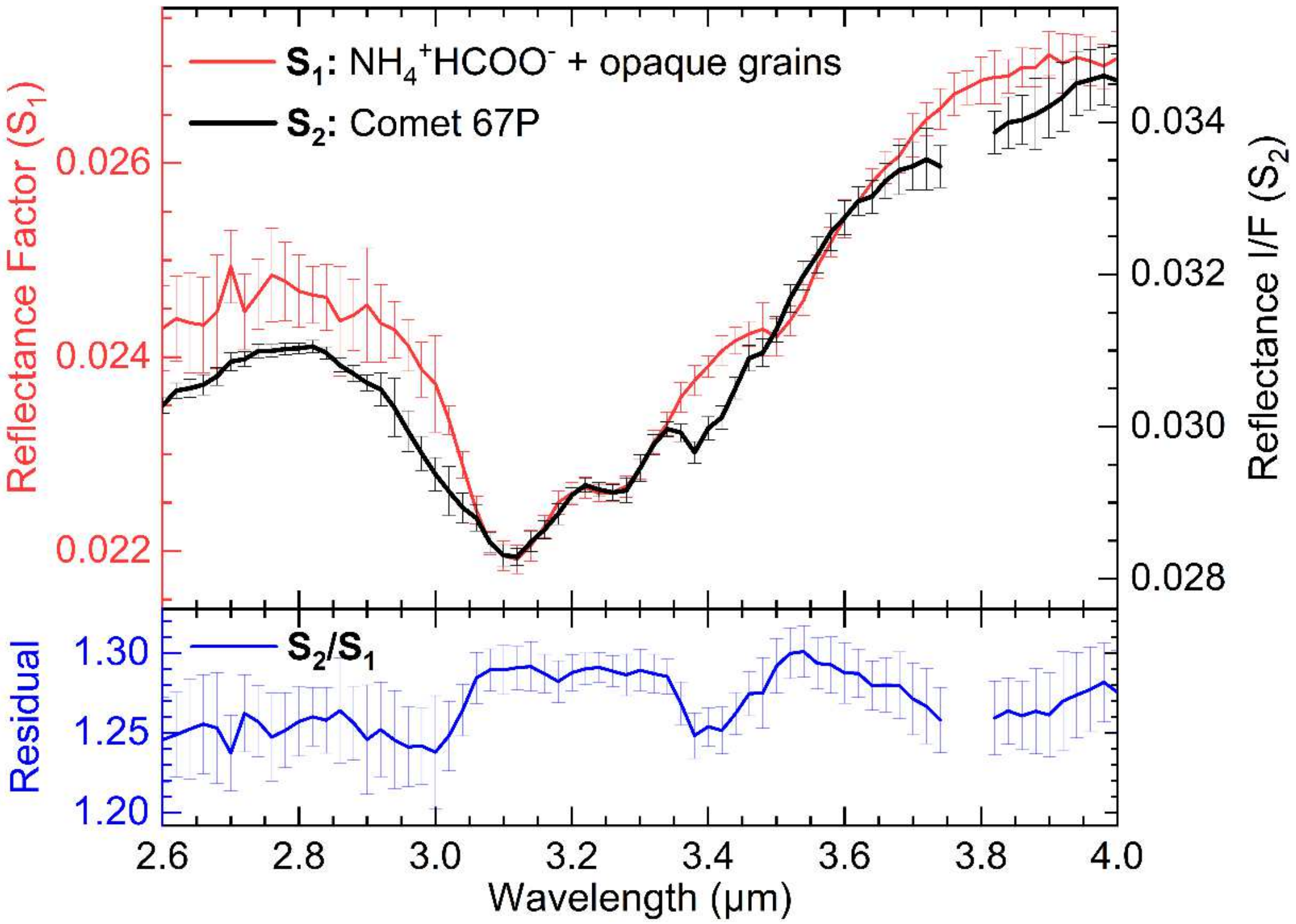

**Fig. S5. Comparison of $NH_4^+$ $HCOO^-$ spectrum with the average spectrum of comet 67P.** Reflectance spectra of the average of the nucleus of comet 67P ($S_1$, black line) and of a sublimate residue containing ≲ 17 wt% ammonium formate mixed with ≳ 83 wt% pyrrhotite grains at 170-200 K ($S_2$, red line). The ratio calculated by dividing the two spectra (blue curve) is also shown. Error bars indicate the ±1σ uncertainty. I/F is the radiance factor.

The similarity of the overall shape of the band and the similarity of the spectra in the range from 3.05 and 3.35 µm on Fig. S5 indicates the presence of ammonium salts on the comet. Differences between these spectra are due to the contribution of other compounds (possibly water ice, carbonaceous compounds) and possibly to different properties of the salts (counter-ions, concentration, mixing etc.) present on the cometary surface. The estimated signal-to-noise ratio is ~70 for the absorption feature at 3.1 µm, ~20 for the feature at 3.3 µm, and ~7 for the feature at 3.4 µm. The spectral ratio between 3.35-3.60 µm indicates the presence of the C−H stretching modes in carbonaceous compounds (*10*). Ammonium salts also exhibit a weaker reflectance minimum centered around 3.50 µm whose identification on the spectrum of comet 67P is impeded by its proximity with the C−H stretching modes and the limited spectral resolution and sampling (respectively of $\Delta\lambda_{resolution} = 0.039$ µm and $\Delta\lambda_{sampling} = 0.020$ µm for the laboratory spectrum and of $\Delta\lambda_{resolution} = 0.015$ µm and $\Delta\lambda_{sampling} = 0.020$ µm for the VIRTIS spectrum after removal of the even bands, as explained above). The position and strength of this ammonium bending mode at around 3.50 µm varies depending on the counter-ions (e.g. it is weak for ammonium citrate and carbamate in Fig. 2) and temperature and/or matrix interaction (Fig. S7A).

A detailed assignment of the bands observed on the cometary spectrum (Figs. 1 and S5) is provided in Table S2.

Several ammonium salts also have absorption bands at shorter wavelengths (at about 1.6-1.7 µm and 2.1-2.3 µm) (*79*). In the comet 67P spectrum and our experimental spectra of sublimate residues, these bands are absent because the salts are mixed with opaque mineral grains, which cause a decrease of the salts' band depths, resulting in a disappearance of the bands at 1.6-1.7 µm and 2.1-2.3 µm because they have lower coefficients of absorption than the ones at 3.1-3.3 µm.

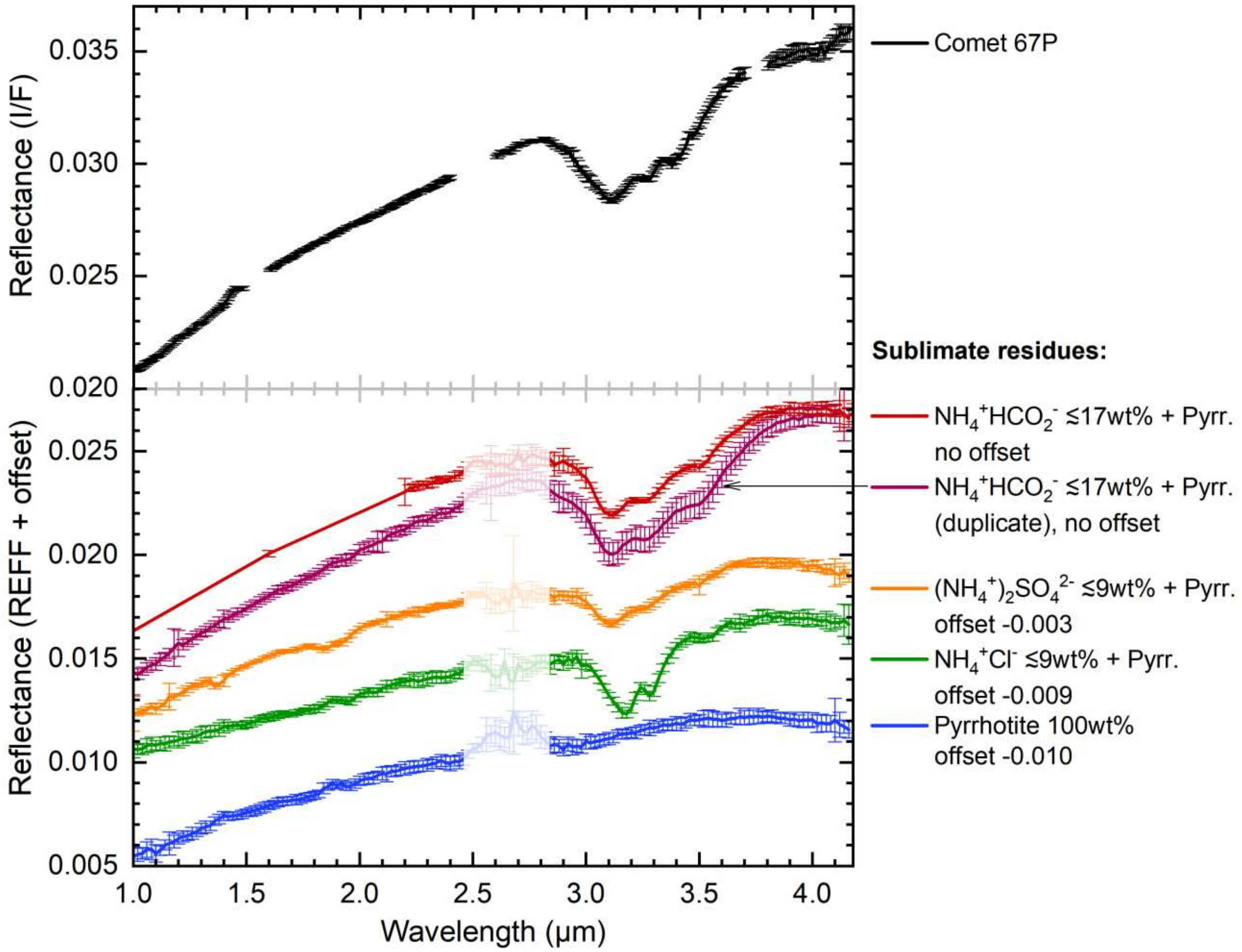

**Fig. S6. Reflectance spectra of sublimate residues containing sub-micrometric-sized pyrrhotite grains pure or mixed with ammonium salts at different concentrations.** These spectra are averages computed from 12 to 30 individual spectra. The red and violet curves show two spectra of sublimate residues containing pyrrhotite grains mixed with ≲ 17 wt% (≲ 43 vol%) ammonium formate ($NH_4^+$ $HCOO^-$), obtained after two duplicated experiments. The error bars reflect the photometric uncertainty and any variation of the samples between individual spectra. Between 2.5 and 2.8 µm, the laboratory spectra are affected by measurement artifacts due to the presence of water vapor in the optical path. Error bars indicate the ±1σ uncertainty. Figure 2 shows these spectra after continuum normalization. I/F is the radiance factor. REFF is the reflectance factor.

The pyrrhotite grains used for the sublimation experiments form a matrix of opaque grains, which are spectrally featureless in the 3-µm region (Fig. 2 and Fig. S6). Mixing ammonium salts with other darkening agents gives similar reflectance spectra. As an example, Fig. S7 shows reflectance spectra of mixtures of ammonium salts and graphite grains mixed in a mortar and measured under ambient conditions.

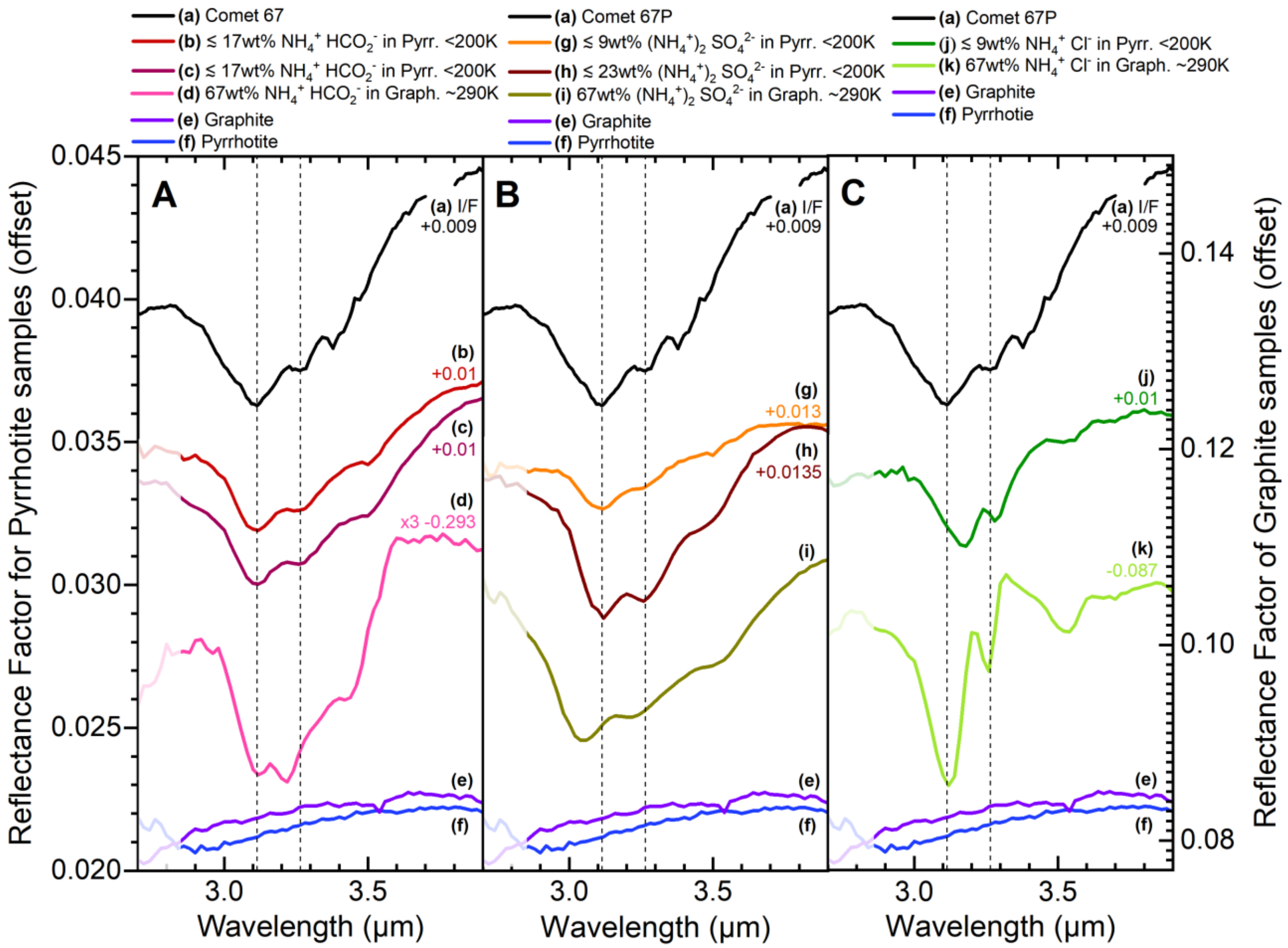

**Fig. S7. Reflectance spectra of ammonium salts mixed with opaque grains of pyrrhotite or graphite.** (**A**) Ammonium formate ($NH_4^+$ $HCOO^-$). (**B**) Ammonium sulfate ($(NH_4^+)_2$ $SO_4^{2-}$). (**C**) Ammonium chloride ($NH_4^+$ $Cl^-$). The mixtures with pyrrhotite (Pyrr.) were obtained by sublimation and were measured under high vacuum (< $10^{-5}$ mbar) and 170-200 K. The mixtures with graphite (Graph.) were obtained by mixing the salt and the graphite grains in a mortar and were measured under ambient temperature and pressure conditions (~290 K). The average reflectance spectrum of the nucleus of comet 67P and the spectra of pure pyrrhotite and graphite grains are also shown for comparison. Between 2.7 and 2.9 µm, the laboratory spectra are affected by measurement artifacts due to the presence of water vapor in the optical path.

The relative amplitudes and the positions of most of the minima of reflectance of the 3.2-µm band of the salts are different in the spectra of the mixtures in graphite compared to the sublimate residues in pyrrhotite. These differences are probably due to the environmental conditions (temperature, difference of interaction with pyrrhotite vs. graphite matrixes etc.) which these ionic species appear to be sensitive to.

Quantity of ammonium salts mixed with the dust of comet 67P

The laboratory reflectance spectrum which is the closest match to comet 67P is obtained with a sublimate residue containing pyrrhotite grains mixed with ≲ 17 wt% ammonium formate ($NH_4^+$ $HCOO^-$) as shown in Fig. 1, Fig. 2 and Fig. S5. However, this upper limit mass fraction of ammonium formate cannot be directly applied to the cometary surface, because this surface is composed of dust grains (made of refractory organic matter and various minerals) much less dense than the pyrrhotite grains present in the sublimate residues. Therefore, for the same mass fraction of salt, the cometary surface layer contains a larger volume of refractory grains than our samples. In order to compare the band depths in our experimental spectra and in the comet spectrum in terms of abundance of salt, we use the volume fraction of the salt in the surface material rather than its mass fraction. Ideally, we would use the cross section of the salt that interacts with the photons, but this value is unknown. Fig. S2 shows that the grains of salt in the sublimate residues are probably sub-micrometric, so we are likely far from the saturation regime of the strong vibration modes of the salts in the 3-μm region.

The upper limit of the volume fraction of salt in the sublimate residue is calculated as follows.

For a two-endmember mixture of A and B, we define the following parameters:

$M$: total mass ($M = M_A + M_B$)
$V$: total volume ($V = V_A + V_B$)
${M_A}^*$: mass fraction of A (${M_A}^* = \frac{M_A}{M}$)
${M_B}^*$: mass fraction of B
${V_A}^*$: volume fraction of A (${V_A}^* = \frac{V_A}{V}$)
${V_B}^*$: volume fraction of B
$\rho_A$: density of A
$\rho_B$: density of B

By definition, we have:

$$\frac{{M_A}^*}{{M_B}^*} = \frac{M_A}{M_B} = \frac{V_A.\rho_A}{V_B.\rho_B} = \frac{{V_A}^*.\rho_A}{{V_B}^*.\rho_B} \quad \text{(S2)}$$

and

$${V_A}^* + {V_B}^* = 1 \quad \text{(S3)}$$
$${M_A}^* + {M_B}^* = 1 \quad \text{(S4)}$$

By combining these equations, we obtain:

$${V_A}^* = \frac{1}{1 + \frac{{M_B}^*.\rho_A}{{M_A}^*.\rho_B}} \quad \text{(S5)}$$

with:

A = ammonium formate ($NH_4^+$ $HCOO^-$), ${M_A}^*$ = 0.17, $\rho_A$ = 1.26 g/cm$^3$,
B = pyrrhotite ($Fe_{1-x}S$, with 0 < x < 0.2), ${M_B}^*$ = 0.83, $\rho_B$ = 4.61 g/cm$^3$.

From Equation S5 we derive an upper limit of the volume fraction of salt of 43 vol% in the sublimate residue.

Fig. S6 shows two spectra of sublimate residues containing pyrrhotite grains mixed with $\lesssim$ 17 wt% ($\lesssim$ 43 vol%) ammonium formate ($NH_4^+$ $HCOO^-$), obtained in duplicated experiments. We calculated the relative band depth at 3.1 µm ($BD$) via

$$BD = 1 - \frac{R}{R_C} \qquad \text{(S6)}$$

with:

$R$ the reflectance value at the wavelength with the minimum reflectance,

$R_C$ the reflectance value of the continuum baseline at this wavelength. The continuum baseline was computed via a spline interpolation (3rd order polynomials) using the reflectance of baseline anchor points typically at 2.20, 2.32, 2.78, 3.96 and 4.12 µm.

The band depths obtained are 0.123 for the comet 67P spectrum and 0.127 and 0.149 for the experimental spectra. Because the band depth of the 3.1 µm absorption band in the spectrum of comet 67P is lower than the band depth of the sublimate residues, the volume fraction of salt on comet 67P should be less than 43 vol%. However, the band depth is controlled by a multitude of parameters in addition to the salt content, such as possible contributions of other surface constituents, grain sizes, porosity and mixing mode of the dust layer, which may differ in our laboratory samples compared to the comet. Even if our samples share some properties with cometary dust (sub-µm grains, porosity, production via crystallization/sublimation of ice), we cannot exclude that the upper limit of ammonium salts on comet 67P is lower or higher than our estimation.

The dark surface material (i.e. dust) of the comet is not composed only of pyrrhotite but of refractory organic matter (ROM) mixed with minerals (MIN). An approximate composition of this material was derived from the analysis of COSIMA data (*52*), resulting in respective abundances of ROM and minerals of $M_{ROM}{}^*$= 0.45 and $M_{MIN}{}^*$= 0.55. Assuming a density of $\rho_{ROM}$= 1 g/cm$^3$ for the ROM and a mean chondritic density of $\rho_{MIN}$= 3.4 g/cm$^3$ for the minerals, we determine:

$$\rho_B = \frac{M_B}{V_B} = \frac{V_{ROM} \cdot \rho_{ROM} + V_{MIN} \cdot \rho_{MIN}}{V_B} = V_{ROM}{}^* \cdot \rho_{ROM} + V_{MIN}{}^* \cdot \rho_{MIN} \qquad \text{(S7)}$$

with:

B = cometary dust,

$V_{ROM}{}^*$ and $V_{MIN}{}^*$ derived from Equation S5, giving $V_{ROM}{}^*$ = 0.74 and $V_{MIN}{}^*$ = 0.26 .

From Equation S7, we find $\rho_B$= 1.63 g/cm$^3$.

Finally, to calculate the mass fraction of salt corresponding to 43 vol% in the cometary dust, we use

$$M_A{}^* = \frac{1}{1 + \frac{V_B{}^* \cdot \rho_B}{V_A{}^* \cdot \rho_A}} \qquad \text{(S8)}$$

with:

A = ammonium formate ($NH_4^+$ $HCOO^-$), $M_A{}^*$ = 0.17, $\rho_A$ = 1.26 g/cm$^3$,

B = cometary dust.

Finally, from Equation S8 we obtain $M_A{}^* = 0.37$

To conclude, on the cometary surface we estimate an upper limit of $\lesssim$ 40 vol% salts, equivalent to $\lesssim$ 40 wt% salts, mixed with the dark surface material.

<u>Calculation of the nitrogen distribution and nitrogen-to-carbon ratio</u>

To calculate the nitrogen budget of comet 67P, we distinguished three main components: i) volatiles; ii) refractory dust grains, made of minerals and ROM; and iii) semi-volatile ammonium salts in the form of ammonium formate ($NH_4^+$ $HCOO^-$, although $NH_4^+$ could have multiple counter-ions on the comet).

The balance of their respective mass fractions can be written:

$$M_V{}^* + M_R{}^* + M_S{}^* = 1 \qquad \text{(S9)}$$

with $M_V{}^*$, $M_R{}^*$ and $M_S{}^*$ the mass fractions of the volatiles, the refractory dust grains and the semi-volatile ammonium salts respectively.

The dust-to-ice mass ratio $R$ is taken as 4 ± 2 (*80*).

$X$ is the mass fraction of ammonium salts with respect to refractory dust:

$$X = \frac{M_S{}^*}{M_S{}^* + M_R{}^*} \qquad \text{(S10)}$$

The terms above can then be rewritten as:

$$M_V{}^* = \frac{1}{R+1} \quad \text{(S11)}; \qquad M_S{}^* = X\frac{R}{R+1} \quad \text{(S12)}; \qquad M_R{}^* = \frac{R}{R+1}(1-X) \quad \text{(S13)}$$

The refractory dust is composed of minerals (MIN) and refractory organic matter (ROM), whose approximate mass ratio was derived from the analysis of COSIMA data (*52*). We assume that COSIMA did not sample the most volatile compounds of the cometary dust, including the ammonium salts that could have sublimated during the pre-analysis storage of the particles at 283 K inside the COSIMA instrument (*47*). Therefore, we have:

$$X = \frac{M_S{}^*}{M_S{}^* + M_R{}^*} \qquad \text{(S14)}$$

$$M_R{}^* = M_{MIN}{}^* + M_{ROM}{}^* \quad \text{(S15)} \qquad \frac{M_{MIN}{}^*}{M_{ROM}{}^*} = \frac{55}{45} \quad \text{(S16)}$$

with $M_{MIN}{}^*$ and $M_{ROM}{}^*$ the mass fractions of the minerals and the ROM respectively.

By combining these equations, we obtain:

$$M_{ROM}{}^* = \frac{M_R{}^*}{1 + \frac{M_{MIN}{}^*}{M_{ROM}{}^*}} = \frac{R(1-X)}{(R+1)(1 + \frac{M_{MIN}{}^*}{M_{ROM}{}^*})} \qquad \text{(S17)}$$

Moles are derived from the mass fractions through:

$$n_V = \frac{M_V{}^*}{M_V} \quad \text{(S18)} ; \qquad n_S = \frac{M_S{}^*}{M_S} \quad \text{(S19)} ; \qquad n_{ROM} = \frac{M_{ROM}{}^*}{M_{ROM}} \quad \text{(S20)}$$

with:

$M_{ROM}$ = 18.29 g/mol, obtained from the elemental composition $HCO_{0.3}N_{0.035}$ estimated from COSIMA data (*47, 52*),

$M_S$ = 63.06 g/mol, assuming ammonium salts are in the form of ammonium formate.

$M_V$ = 18.96 g/mol, assuming the volatiles are composed of the sixteen most abundant molecules detected in the summer hemisphere by the ROSINA instrument onboard Rosetta (*16, 50*): $H_2O$, CO, $CO_2$, $CH_4$, $C_2H_2$, $C_2H_6$, $H_2CO$, HCOOH, $CH_2OHCH_2OH$, $HCOOCH_3$, $NH_3$, HCN, $N_2$, $CH_3CHO$, HNCO, $CH_3CN$. We assume that the bulk comet has the same composition than the coma, which might not be valid for super-volatiles (*81*). Nevertheless, the contribution of ices to the nitrogen budget is small, so this assumption should have a negligible impact.

Finally, the total number of carbon ($n^C$) and nitrogen ($n^N$) atoms is calculated for each reservoir:

$$n_V^C = f_V^C . n_V \quad \text{(S21)} ; \qquad n_S^C = f_S^C . n_S \quad \text{(S22)} ; \qquad n_{ROM}^C = f_{ROM}^C . n_{ROM} \quad \text{(S23)}$$

$$n_V^N = f_V^N . n_V \quad \text{(S24)} ; \qquad n_S^N = f_S^N . n_S \quad \text{(S25)} ; \qquad n_{ROM}^N = f_{ROM}^N . n_{ROM} \quad \text{(S26)}$$

with $f^C$ and $f^N$ respectively the fractions of carbon and nitrogen atoms per mole of each reservoir.

The $N/C$ ratio is calculated via:

$$N/C = \frac{n_V^N + n_S^N + n_{ROM}^N}{n_V^C + n_S^C + n_{ROM}^C} \qquad \text{(S27)}$$

The results when varying $X$ from 0 to 40 wt% are shown in Figure 4 for pure ammonium formate, and in Figure 5 for different ammonium salts.

The volatiles contribute little to the nitrogen budget (less than 1% N), whatever the dust-to-ice mass ratio within the range from 2 to 6. Fig. 4 shows that the main nitrogen reservoirs are then located in the ROM (presumably in aromatic rings and cyanide chemical groups) and in semi-volatile ammonium ions with their (potentially nitrogen-bearing) counter-ions. Fig. 5 shows ammonium salts may considerably increase the N/C ratio of the whole comet, from 0.035 possibly up to values close to the solar value of 0.29 ± 0.12 (*48*).

These calculations rely on the single value of 0.035 for the N/C ratio in the ROM, which is based on the analysis of a few tens of grains (*47*). We can further test the sensitivity of the nitrogen budget to this parameter by considering a N/C of 0.120, consistent with reported values for N-rich IDPs and micrometeorites of plausible cometary origin (*82*). The distribution of the nitrogen resulting from this calculation is shown in Fig. S8 and the N/C ratio in Fig. S9. Such high N/C values in the ROM were only measured in very rare ultra-carbonaceous Antarctic micro-meteorites (*82*), and such particles are unlikely to represent the bulk N/C of the refractory carbon of a whole comet.

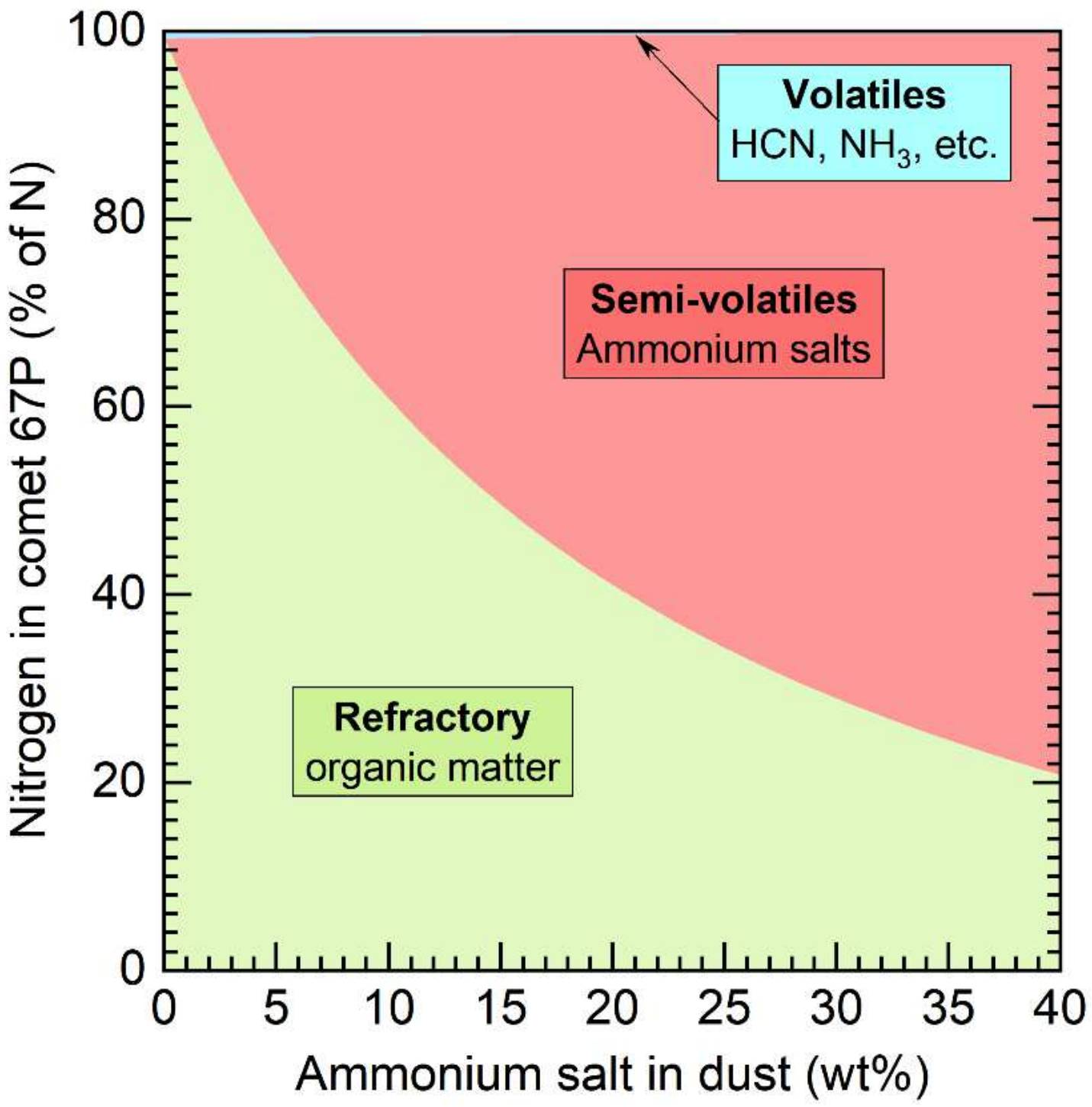

**Fig. S8. Distribution of nitrogen in comet 67P.** Same as Figure 4, but for a refractory organic matter having a $(N/C)_{ROM}$ ratio of 0.120.

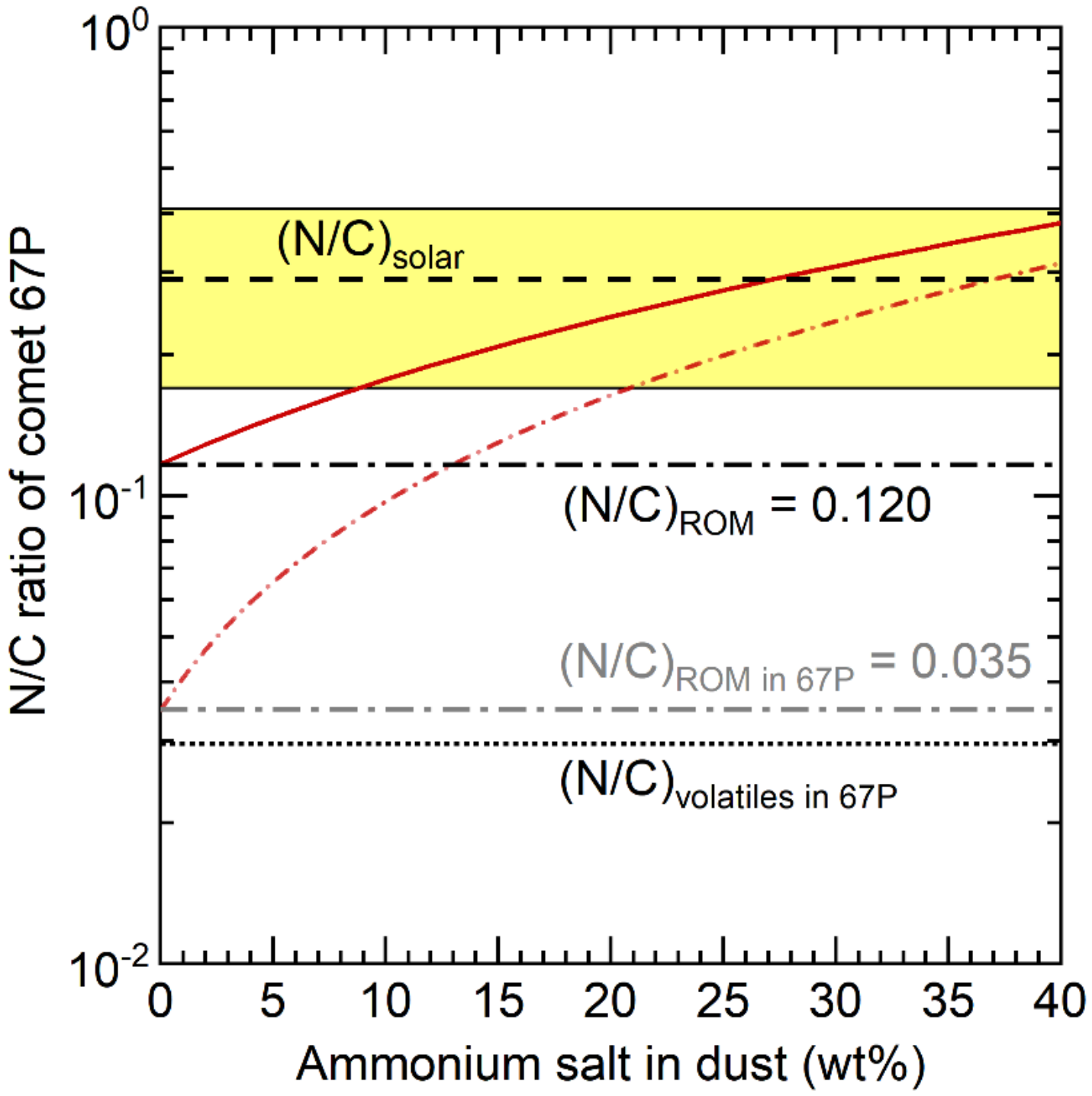

**Fig. S9. Nitrogen-to-carbon ratio in comet 67P.** Same as Figure 5, but for $(N/C)_{ROM}$ = 0.120 (red line). The case with $(N/C)_{ROM}$ = 0.035 (red short dashed-dotted line) is also shown. The ammonium salt is assumed to be made only of ammonium formate ($NH_4^+$ $HCOO^-$) in this example.

| Observed positions of absorption maxima (±0.01 µm) | | | | | | | | | Assignment |
|---|---|---|---|---|---|---|---|---|---|
| **Comet 67P** | **$NH_4^+$ $HCOO^-$** | | **$(NH_4^+)_2$ $SO_4^{2-}$** | | **$NH_4^+$ $Cl^-$** | | **$NH_4^+$ $CN^-$** | **$NH_4^+$ $OCN^-$** | |
| Ref. [a, b] | This work* | Ref. [c] | This work* | Ref. [d, e] | This work* | Ref. [f, g] | Ref. [h] | Ref. [i, k] | |
| | | | | 3.03 | | | | | $NH_4^+$ asym. stretching $\nu_3$ [d, e] |
| **3.11** | 3.12 | 3.13 | 3.12 | 3.10-3.11 | | | | 3.12 [k] | $NH_4^+$ comb. $\nu_1$ (sym.) + $\nu_5$ (lattice) [j] |
| | | | | | | | 3.15 | 3.15-3.16 | $NH_4^+$ asym. stretching $\nu_3$ and/or $\nu_2$ (sym. bending) + $\nu_4$ (asym. bending) [j, h] |
| | | | | | 3.18 | 3.17-3.20 | | | $NH_4^+$ asym. stretching $\nu_3$ [j] |
| **3.26** | 3.26 | 3.25 | 3.26 | 3.26-3.28 | 3.28 | 3.28 | | | $NH_4^+$ asym. stretching $\nu_3$ and/or $\nu_2$ (sym. bending) + $\nu_4$ (asym. bending) [j] |
| | | | | | | | 3.29 | 3.29 | $NH_4^+$ asym. stretching $\nu_3$ and/or $\nu_2$ (sym. bending) + $\nu_4$ (asym. bending) [h] |
| | | 3.32 | | | | | | | N−H stretching in $NH_4^+$ [c] |
| **3.38** | | | | | | | | | C−H stretching [a] |
| **3.42** | | | | | | | | | C−H stretching [a] |
| **3.47** | | 3.48 | | | | | | | C−H stretching [a]<br>N−H stretching in $NH_4^+$ [c] |
| | 3.5 | | 3.5 | 3.50-3.52 | 3.52-3.56 | 3.52 | 3.49 | 3.50-3.51 | $NH_4^+$ asym. bending overtone $2\nu_4$ [j, h] |
| | | 3.56 | | | | | | | C−H stretching in $HCOO^-$ [c] |
| | | 3.69 | | | | | | | 2 × C−H in plane bending in $HCOO^-$ [c] |

**Table S2. Observed absorption bands in the spectra of the nucleus of comet 67P and of some ammonium salts, with their assignments.** The absorption bands at 3.11 and 3.26 µm in the spectrum of the comet 67P are assigned to fundamental and/or combination modes of N−H vibrations in $NH_4^+$ in ammonium salts, such as ammonium formate and sulfate. Ammonium cyanide and cyanate (although not measured under the same conditions) could also partly contribute to the broad absorption band in the 67P spectrum, but their band centers appear shifted toward longer wavelengths. Ammonium chloride appears as a less likely contributor. The absorptions at 3.38, 3.42 and 3.47 µm in the comet spectrum are attributed to C−H stretching vibrations of organic compounds (*10*). The identification on the spectrum of comet 67P of a weaker N−H mode of ammonium salts centered around 3.50 µm is impeded by the proximity of the C−H stretching modes and the limited spectral resolution and sampling (respectively of 0.039 µm and 0.020 µm for the laboratory spectrum and of 0.015 µm and 0.020 µm for the VIRTIS spectrum). Complementary spectroscopic data on ammonium nitrates, phosphates, carbonates and oxalates have been published (*79*).

[*] Reflectance spectra of the ammonium salts in sublimate residues with pyrrhotite sub-micrometric-sized grains at 170-200 K and under high vacuum ($< 10^{-5}$ mbar). The fundamental vibrations of $NH_4^+$ listed here are $\nu_1$, the symmetric stretching mode, $\nu_2$, the symmetric bending mode, $\nu_3$, the asymmetric stretching mode, $\nu_4$, the asymmetric bending mode, and $\nu_5$ is a lattice mode.

[a] Average reflectance spectrum of the surface of comet 67P obtained from our VIRTIS-M data. The absorption features observed at 3.38, 3.42 and 3.47 µm in this spectrum are attributed to C−H stretching from organic matter (*10*).

[b] Average reflectance spectrum of the surface of comet 67P obtained from VIRTIS-M data (previous calibration) (*5*)

[c] Transmittance spectrum of $NH_4^+$ $HCOO^-$ at 298 K (*83*)

[d] Transmittance spectrum of $(NH_4^+)_2$ $SO_4^{2-}$ at ambient temperature and pressure conditions (*84*)

[e] Transmittance spectra of $(NH_4^+)_2$ $SO_4^{2-}$ from 288 to 183 K in (*85*). The relative amplitude of the absorptions at 3.1 and 3.3 µm change dramatically with the temperature (*85*).

[f] Reflectance spectrum of $NH_4^+$ $Cl^-$ at 123 K (*86*)

[g] Transmittance spectrum of $NH_4^+$ $Cl^-$ at 250 K (*87*)

[h] Transmittance spectrum of $NH_4^+$ $CN^-$ at 125 K (*88*)

[i] Transmittance spectrum of $NH_4^+$ $OCN^-$ at 250 K (*87*)

[j] Assignment according to (*89*). The intense absorptions of ammonium salts in the 3.2-µm region are proposed to be due to Fermi resonance between the fundamental $\nu_3$ and the combination $\nu_2 + \nu_4$ (*86*, *89*).

[k] Transmittance spectrum of $NH_4^+$ $OCN^-$ at 160 K (*21*)